\documentclass[11pt,a4paper]{article}
\usepackage[T1]{fontenc}
\usepackage[utf8]{inputenc}
\usepackage{lmodern,microtype}
\usepackage[top=25mm,bottom=27mm,left=27mm,right=27mm]{geometry}
\usepackage{amsmath,amssymb,amsthm,mathtools,bm}
\usepackage{graphicx,booktabs,array,tabularx,enumitem,placeins}
\usepackage[numbers,sort&compress,square]{natbib}
\usepackage{xcolor}
\usepackage[colorlinks=true,linkcolor=blue!45!black,citecolor=blue!45!black,urlcolor=blue!45!black]{hyperref}
\hypersetup{pdftitle={Recovering classical gauge orbits from quantum Dirac constraints},pdfauthor={Alejandro Corichi}}
\graphicspath{{figures/}}
\renewcommand{\thesection}{\Roman{section}}

\makeatletter
\renewcommand{\p@subsection}{\thesection\,}

\@addtoreset{equation}{section}
\makeatother

\usepackage{etoolbox}
\makeatletter
\patchcmd{\l@section}{1.5em}{2.6em}{}{}
\renewcommand*\l@subsection{\@dottedtocline{2}{2.6em}{2.4em}}
\makeatother
\allowdisplaybreaks[1]
\setlist{itemsep=2pt,topsep=4pt}

\theoremstyle{plain}
\newtheorem{theorem}{Theorem}
\newtheorem{proposition}[theorem]{Proposition}
\newtheorem{lemma}[theorem]{Lemma}
\newtheorem{corollary}[theorem]{Corollary}
\newtheorem{conjecture}[theorem]{Conjecture}
\theoremstyle{remark}
\newtheorem{remark}[theorem]{Remark}

\newcommand{\C}{\hat C}
\newcommand{\R}{\mathbb R}
\newcommand{\Cx}{\mathbb C}
\newcommand{\Z}{\mathbb Z}
\newcommand{\PP}{\mathbb P}
\newcommand{\Hil}{\mathcal H}
\newcommand{\Pphy}{\mathcal P_{\mathrm{phy}}}
\newcommand{\GC}{G_{\Cx}}
\newcommand{\g}{\mathfrak g}
\newcommand{\dd}{\mathrm d}
\newcommand{\Var}{\operatorname{Var}}
\newcommand{\Cov}{\operatorname{Cov}}
\newcommand{\Vol}{\operatorname{Vol}}
\newcommand{\vol}{\operatorname{vol}}
\renewcommand{\Re}{\operatorname{Re}}
\renewcommand{\Im}{\operatorname{Im}}
\newcommand{\red}{\mathrm{red}}
\newcommand{\GB}{\mathrm{GB}}
\newcommand{\ket}[1]{|#1\rangle}

\newcommand{\cs}[1]{|#1)}

\begin{document}

\begin{center}
{\Large\bfseries Recovering classical gauge orbits\\[2pt] from quantum Dirac constraints\par}
\vspace{12pt}
{\large Alejandro Corichi\par}
\vspace{4pt}
{\small\itshape Centro de Ciencias Matem\'aticas, Universidad Nacional Aut\'onoma de M\'exico,\\ UNAM-Campus Morelia, A.\ Postal 61-3, Morelia, Michoac\'an 58190, M\'exico\par}
\vspace{2pt}
{\small\texttt{corichi@matmor.unam.mx}\par}
\end{center}

\begin{abstract}
\noindent
Classical reduction of a system with first-class constraints takes two steps: one restricts to the constraint surface and then identifies the points of each gauge orbit. Dirac quantization takes one step, and, as shown in earlier work, the resulting space of physical rays is a K\"ahler manifold with no gauge directions left. Where, then, does the quantum theory encode the classical gauge structure? We find it in two complementary places. The first is motion: under explicit correspondence conditions, the velocities that constraint unitaries impart to localized kinematical packets span the null directions of the constraint surface. The second is projection. For any constraints with a common kernel, closing or not, the physical weight $F_\hbar$ of a state, the probability that it passes the test ``is it physical?'', is invariant under the real gauge flows and changes along the imaginary ones at a rate set by the constraint values. For a compact gauge group $-\log F_\hbar$ is a Kempf--Ness function, so along each complexified gauge orbit $F_\hbar$ is maximal exactly where all constraint expectations vanish. With coherent states for linear actions, or heat-kernel packets on cotangent bundles of compact groups, this recovers the constraint surface, its gauge orbits and the reduced K\"ahler space at every $\hbar$. Resolving a gauge orbit forces $F_\hbar$ to vanish; our packets attain this bound, with $F_\hbar$ equal to $(4\pi\hbar)^{k/2}$ over the volume of the $k$-dimensional orbit up to $O(\hbar)$, a law we conjecture to hold in general. Examples include structure functions, stabilizers, continuous spectrum and an anomaly.
\end{abstract}


\tableofcontents

\section{Introduction}\label{sec:intro}

Most theories of fundamental interest are subject to constraints among their canonical variables when they are written in Hamiltonian form. Gauge theories are the familiar examples, and general relativity is the extreme one: on a compact Cauchy surface its Hamiltonian is itself a combination of constraints, so that even time evolution is, in a sense, generated by gauge transformations. The classical treatment of such systems is by now well understood in geometric terms~\cite{dirac,ht}. One first restricts attention to the constraint surface $\Sigma$, on which the first-class constraints $C_a$ vanish. The symplectic form pulled back to $\Sigma$ is degenerate, its null directions are generated by the constraints themselves, and points on the same null leaf, a gauge orbit, describe the same physical state. The second step is to take the quotient of $\Sigma$ by the gauge orbits, which yields the reduced phase space $\Gamma_\red$. Classical reduction therefore takes two steps: restriction, and then quotient.

In Dirac's approach to the quantum theory the prescription is shorter~\cite{dirac}. The constraints are promoted to operators $\C_a$ on a kinematical Hilbert space, and physical states are required to satisfy $\C_a\psi=0$. There is a single quantum condition, and it is often said that this one step accomplishes both classical ones. In earlier work~\cite{corichi2008} we made this statement precise using the geometric formulation of quantum mechanics~\cite{kibble,ashtekarschilling}, in which the space of rays is a K\"ahler manifold and unitary transformations are its canonical transformations. The Dirac condition imposes the vanishing of the constraint expectation values, the quantum analogue of $C_a=0$, and, in addition, the vanishing of the gauge vector fields themselves. The space of physical rays $\Pphy$ is then a totally geodesic K\"ahler submanifold, its symplectic form is nondegenerate, and there are no gauge directions left to quotient by. The quantum theory does not identify the points of a gauge orbit; it retains only those states whose orbit is a single point.

This answer raises the question that is the starting point of the present paper. If the physical state space carries no trace of the gauge orbits, where in the quantum theory is the classical gauge structure encoded? The question is not academic. The classical limit of a constrained quantum theory should reproduce not only the reduced phase space but also the unreduced description in which the theory is usually formulated, with its constraint surface and its gauge orbits. It is in that description that we write the Einstein equations or the Gauss law, and in which semiclassical states are usually constructed~\cite{abc2005,thiemannGCS}. In quantum gravity the issue is sharper still. In several approaches one has, or hopes to have, direct access to the physical Hilbert space, for instance in deparametrized models or in loop quantum cosmology with a massless scalar field~\cite{aps2006}. One would then like to know which features of the classical gauge structure can be recovered from the physical theory, and which require additional, kinematical, input.

A simple example shows where to look. Consider a particle on a circle constrained to have zero angular momentum, $\hat p\psi=0$. The constant wavefunction solves the constraint exactly, and rotating it reveals nothing: it has no angular location. A kinematical packet with mean momentum zero, by contrast, can be sharply localized about an angle. If we rotate this packet its center moves; if we project it onto the zero mode the center disappears. Following the center tells us which classical descriptions are related by the gauge; projecting tells us what is left of their distinction in the physical state. There is a second, less obvious, lesson. Push the same packet in imaginary time, $\psi\mapsto e^{\tau\hat p/\hbar}\psi$. Its mean momentum moves off zero, and yet its projection is unchanged. The fibers of the projection onto physical states therefore contain \emph{complexified} gauge orbits, not only gauge orbits. What singles out the classical surface $p=0$ inside such a fiber is the weight $F_\hbar=\|P\psi\|^2/\|\psi\|^2$ of the physical component, that is, the probability that the kinematical state passes the test ``is it physical?''. It is largest exactly where $\langle\hat p\rangle=0$.

In this paper we develop these observations into two complementary readings of the classical gauge geometry.

\emph{Motion} (Sec.~\ref{sec:motion}). The individual constraint unitaries move localized kinematical packets, and kinematical probes read the velocities of the packet labels. Under explicit correspondence and regularity conditions these velocities span the kernel of the classical two-form restricted to $\Sigma$. This test is what licenses calling the observed motion gauge: many unitaries move a packet, but not every motion relates classically equivalent descriptions.

\emph{Projection} (Secs.~\ref{sec:exact}, \ref{sec:projection} and \ref{sec:landscape}). This reading uses only the physical projector and the weight $F_\hbar$. Exact identities hold for any constraints with a common kernel, with no group and no closure of the algebra. For a compact gauge group they integrate to the Kempf--Ness theorem, read through the physical weight. With a packet family on which the complexified group acts, $\Sigma$, its gauge orbits and the reduced space are recovered exactly, at every $\hbar$. The classical model for this reading is the holomorphic description of K\"ahler reduction, which we recall in Sec.~\ref{sec:classical}.

Both readings need a semiclassical interpretation of the kinematical theory. We never infer an unreduced classical description from the geometry of $\Pphy$ alone, and Sec.~\ref{sec:torus} shows why one cannot: theories whose gauge leaves have different dimensions can have isometric physical ray spaces. What the projection reading adds is that the extra structure required is small. A packet family covariant under the complexified gauge group suffices.

The two readings also meet quantitatively. Resolving many points of one orbit forces $F_\hbar$ to vanish (Theorem~\ref{thm:resolved}). For packets calibrated by the K\"ahler metric this bound is attained: on $\Sigma$, $F_\hbar$ is one over the number of Planck cells of size $(4\pi\hbar)^{k/2}$ that cover the $k$-dimensional gauge orbit (Proposition~\ref{prop:volume} and Fig.~\ref{fig:landscape}).

The main results of the paper are the following.
\begin{enumerate}[label=(\roman*)]
\item \emph{Exact identities for the physical weight} (Proposition~\ref{prop:identities}). For any self-adjoint constraints with a common kernel, with no group and no closure, the real gauge flows preserve $F_\hbar$, and along the imaginary ones $\log F_\hbar$ changes at the rate $-2c_a/\hbar$. The quantum constraint surface is therefore the set where $F_\hbar$ is stationary along the imaginary gauge directions, a characterization that uses no master function.
\item \emph{Dirac's classification at finite $\hbar$} (Proposition~\ref{prop:classification}). Structure constants, central anomalies and structure functions are distinguished by the coisotropy of the zero set of the constraint expectations. An explicit model shows that this zero set, and its coisotropy, depend on how the constraints are written, while $F_\hbar$ does not.
\item \emph{The motion reading} (Proposition~\ref{prop:local}), with its hypotheses stated explicitly and tested in the examples.
\item \emph{The Kempf--Ness theorem for the physical weight} (Theorem~\ref{thm:KN}). For a compact group $-\log F_\hbar$ is, up to a constant on each complexified orbit, a Kempf--Ness function. The rays that K\"ahler reduction selects are precisely those that Dirac projection loses least, and near them $F_\hbar$ is Gaussian in the constraint values.
\item \emph{Recovery from $\Pphy$} (Theorems~\ref{thm:linear} and~\ref{thm:hall}). For linear actions with coherent states, and for heat-kernel packets on $T^*G$, the constraint surface, its gauge orbits and the reduced K\"ahler space with its potential are determined by $\Pphy$ and the packet family alone. The ingredients are classical; the statement that the physical weight alone selects $\Sigma$ within each projection fiber appears to be new.
\item \emph{A resolution bound} (Theorem~\ref{thm:resolved} and Proposition~\ref{prop:haar}). Resolving a gauge orbit forces $F_\hbar$ to vanish, for any constraint words and with no group; for compact groups the phases of the overlaps carry the constraint values.
\item \emph{The orbit-volume law} (Proposition~\ref{prop:volume} and Conjecture~\ref{conj:intrinsic}). On $\Sigma$, $F_\hbar=(4\pi\hbar)^{k/2}/\Vol_g$ up to $O(\hbar)$. For coherent states this is the Bargmann--Fock counterpart of known asymptotics of invariant Szeg\H{o} and Bergman kernels~\cite{paoletti2005,paoletti2008,mazhang}; here it is extended to heat-kernel packets and explained in terms of calibration.
\end{enumerate}

Let us place these results in context. The geometric formulation of constrained quantum systems used here was developed in~\cite{corichi2008}, whose results we recall where needed. The same question was approached through master constraints in~\cite{corichiMC}, where the gauge directions are read from the linearized master flow, and the geometry of group averaging is studied in~\cite{corichiGA}. Using packets complements the construction of physical semiclassical states by averaging kinematical states~\cite{abc2005}: there the question is which physical state results from averaging, here it is what motion can be recovered before the gauge label disappears. Coherent-state matrix elements of the physical projector are the basic object of Klauder's projection-operator approach to constraints~\cite{klauder1997,klauder2001}, where they serve to build reduced path integrals; we use them instead to locate $\Sigma$. For compact groups, the convexity behind Sec.~\ref{sec:projection} is that of the Kempf--Ness function~\cite{kempfness,woodward}, which also underlies the moment-map approach to quantum marginals and entanglement classes~\cite{walter2013,burgisser2019}; that invariant states concentrate near the zeros of the moment map is the content of the Guillemin--Sternberg theorem~\cite{guilleminsternberg}. On the zero set, the diagonal of invariant Szeg\H{o} and Bergman kernels involves the volume of the gauge orbits~\cite{paoletti2005,paoletti2008,mazhang}, the same volume that obstructs asymptotic unitarity of the Guillemin--Sternberg map~\cite{hallkirwin}; projection asymptotics for circle actions are studied in~\cite{zelditchzhou}. For SU(2) closure constraints, related coherent-state constructions are due to Conrady and Freidel~\cite{conradyfreidel} and to Bianchi, Don\'a and Speziale~\cite{bds}; Sec.~\ref{sec:discussion} lists what must be compared.

The structure of the paper is as follows. In Sec.~\ref{sec:exact} we recall classical reduction, with emphasis on its holomorphic description for K\"ahler phase spaces, fix our conventions for geometric quantum mechanics and prove the identities for the physical weight that need no group. In Sec.~\ref{sec:dirac} we read Dirac's classification off the state space at finite $\hbar$ and show that it depends on the presentation of the constraints. Section~\ref{sec:motion} develops the motion reading and Sec.~\ref{sec:projection} the projection reading, which is the core of the paper. In Sec.~\ref{sec:landscape} we study the landscape of the physical weight, which connects the two readings quantitatively. Section~\ref{sec:examples} works out the examples, and we end with a discussion in Sec.~\ref{sec:discussion}. Appendix~\ref{app:checks} records the numerical checks.

\section{Classical reduction and exact quantum imposition}\label{sec:exact}

In this section we collect the classical and quantum ingredients that we shall need. This section has three parts. In the first one we recall classical reduction in symplectic language, with emphasis on the role of the complexified gauge group when the phase space is K\"ahler; this classical picture is the one that the projection reading of Sec.~\ref{sec:projection} lifts to the quantum theory. In the second part we fix our conventions for geometric quantum mechanics and recall from~\cite{corichi2008} what exact imposition of the constraints means on the space of rays. In the third part we prove exact identities for the physical weight. Along the way we shall see that each Dirac condition is two conditions on a ray, and that the physical weight $F_\hbar$ obeys exact identities along the real and imaginary constraint flows. Neither statement needs a group, closure of the constraint algebra or a master function.

\subsection{Classical reduction in two steps}\label{sec:classical}

Let $(\Gamma,\omega)$ be a phase space and $C_a$, $a=1,\dots,N$, a set of constraints, with constraint surface $\Sigma=\{C_a=0\}$. We use the conventions $\iota_{X_f}\omega=\dd f$ and $\{f,k\}=\dd f(X_k)$. We assume that $\Sigma$ is regular, that is, that the differentials $\dd C_a$ are independent on it, and that the constraints are first class, $\{C_a,C_b\}\approx0$, where $\approx$ denotes equality on $\Sigma$. Then the Hamiltonian vector fields $X_{C_a}$ are tangent to $\Sigma$, and for every vector $V$ tangent to $\Sigma$ one has $\omega(X_{C_a},V)=\dd C_a(V)=0$. Since the $\dd C_a$ span the conormal space of $\Sigma$, the $X_{C_a}$ span the kernel of the pullback $\omega|_\Sigma$: the surface $\Sigma$ is coisotropic, its null distribution integrates to the gauge orbits, and the reduced phase space is the space of orbits $\Gamma_\red=\Sigma/\!\sim$, when this is a manifold. Physical observables are the functions constant along the gauge orbits. They are precisely the functions on $\Gamma_\red$, which inherits a nondegenerate symplectic form $\omega_\red$. When the constraints have constant structure coefficients they generate a Hamiltonian action of a Lie group $G$ with moment map $\mu=(C_a):\Gamma\to\g^*$, and $\Gamma_\red=\mu^{-1}(0)/G$ is the Marsden--Weinstein quotient~\cite{marsdenweinstein}.

Suppose now that $\Gamma$ is K\"ahler, with complex structure $J$ and metric $g=\omega(\cdot,J\cdot)$, that $G$ is compact and acts by K\"ahler isometries, and that the action extends to a holomorphic action of the complexified group $\GC$. The gradient of each constraint is then $\nabla C_a=JX_{C_a}$. These \emph{imaginary} gauge directions play no role in the symplectic description, but they are transverse to $\Sigma$, and since $J$ is an isometry the Gram matrix $g(X_{C_a},X_{C_b})=g(\nabla C_a,\nabla C_b)=\dd C_a(\nabla C_b)$ measures at once the speeds along the gauge orbits and the rates at which the imaginary flows leave $\Sigma$. For a positive quadratic master function, the linearized master flow on $\Sigma$ maps the imaginary directions onto the real ones through the same matrix, weighted by the master weights~\cite{corichiMC}. The orbits of $\GC$ thus combine the gauge orbits, which lie in $\Sigma$, with directions transverse to it, and the second step of the reduction can be carried out holomorphically. This is the content of the Kempf--Ness theorem~\cite{kempfness,woodward,git}, which we state for linear actions, the case we shall use most.

Let $\Gamma=\Cx^n$ with $\omega=i\sum\dd z\wedge\dd\bar z$, let $u$ be a unitary representation of $G$ on $\Cx^n$, and let $\mu_\xi(z)=i\,\bar z\cdot\dd u(\xi)z$ be its moment map. Consider the function $h\mapsto|u(h)z|^2$ on $\GC$. It is invariant under $h\mapsto kh$ for $k\in G$, it is convex along the curves $\tau\mapsto\exp(i\tau\xi)h$, and its derivative along them is the moment map:
\begin{equation}\label{eq:classKN}
\frac{\dd}{\dd\tau}\,\big|u(\exp(i\tau\xi))z\big|^2=2\,\mu_\xi\big(u(\exp(i\tau\xi))z\big).
\end{equation}
Hence a point of the orbit $\GC\cdot z$ is critical for the norm exactly where $\mu=0$, the critical points are minima and form a single $G$-orbit, and $\GC\cdot z$ meets $\mu^{-1}(0)$ if and only if it is closed. The natural map $\Sigma/G\to\Gamma^{\mathrm{ps}}/\GC$, where the \emph{polystable} set $\Gamma^{\mathrm{ps}}$ consists of the points with closed complexified orbit, is therefore a bijection. The K\"ahler structure of the quotient is also encoded holomorphically: the minimum of $|u(h)z|^2$ over $h\in\GC$, which equals $|z|^2$ on $\Sigma$, is a K\"ahler potential for $\omega_\red$~\cite{hklr}.

Two features of this classical picture will reappear in the quantum theory. First, the imaginary gauge directions are what allow one to locate $\Sigma$ within a complexified orbit: $\Sigma$ is where the squared norm, whose logarithm is the Kempf--Ness function~\cite{woodward}, is stationary along them. Second, the fibers of the holomorphic description of the quotient are the complexified orbits, not the gauge orbits. In Sec.~\ref{sec:projection} we shall see that Dirac projection realizes both features: its fibers contain the complexified gauge orbits, and the role of the squared norm is played by $-\hbar\log F_\hbar$.

\subsection{Geometric quantum mechanics and exact imposition}\label{sec:gqm}

Let us now turn to the quantum theory. In the geometric formulation of quantum mechanics~\cite{kibble,ashtekarschilling} the space of rays $\PP(\Hil)$ of the kinematical Hilbert space is a K\"ahler manifold. Its symplectic form implements unitary transformations as canonical transformations, while its metric encodes uncertainties and transition probabilities. In~\cite{corichi2008} we used this structure to describe Dirac quantization in the same geometric language as its classical counterpart; here we recall what we need.

\paragraph{Conventions.} The inner product is linear in its second slot. A tangent vector at a normalized ray $[\psi]$ of $\PP(\Hil)$ is represented by $v\perp\psi$. The K\"ahler structure~\cite{ashtekarschilling,corichi2008} reads, in our normalization,
\begin{equation}\label{eq:kahler}
g_\hbar(u,v)=2\hbar\,\Re\langle u|v\rangle,\qquad \Omega_\hbar(u,v)=2\hbar\,\Im\langle u|v\rangle,\qquad Jv=iv.
\end{equation}
We use the conventions of Sec.~\ref{sec:classical}, $\iota_{X_f}\Omega_\hbar=\dd f$ and $\{f,k\}=\dd f(X_k)$. The constraints $\C_a$ are self-adjoint; for unbounded generators, differentiations and commutators are taken on the stated domains or on an invariant smooth core. At $\psi$ in the domain of $\C_a$ write $c_a=\langle\C_a\rangle_\psi$ and $d_a=(\C_a-c_a)\psi$. The Hamiltonian field, the gradient and the squared residual are
\begin{equation}\label{eq:fields}
X_{c_a}=-\tfrac{i}{\hbar}\,d_a,\qquad \nabla c_a=JX_{c_a}=\tfrac{1}{\hbar}\,d_a,\qquad \|\C_a\psi\|^2=c_a^2+\tfrac{\hbar}{2}\,g_\hbar(X_{c_a},X_{c_a}).
\end{equation}
$X_{c_a}$ generates the unitary flow $[e^{-it\C_a/\hbar}\psi]$ and $\nabla c_a$ generates the normalized imaginary flow $[e^{\tau\C_a/\hbar}\psi]$. We call $X_{c_a}$ the real and $\nabla c_a$ the imaginary gauge directions. The last identity in~\eqref{eq:fields} is the one-constraint case of the decomposition of the master constraint into the squared moment map and the squared orbit speed~\cite{corichi2008}. The following reformulation of a single Dirac condition is also in~\cite{corichi2008}; we include its one-line proof.

\begin{lemma}[one Dirac condition]\label{lem:dirac}
For normalized $\psi$ in the domain of $\C$, $\C\psi=0$ if and only if $c=0$ and $\dd c=0$ at $[\psi]$.
\end{lemma}
\begin{proof}
$X_c$ vanishes exactly when $\psi$ is an eigenvector of $\C$ with eigenvalue $c$. Since $\Omega_\hbar$ is nondegenerate, $\dd c=\Omega_\hbar(X_c,\cdot)$ vanishes exactly when $X_c$ does.
\end{proof}

So exact imposition means zero mean and zero orbit velocity; zero velocity alone only identifies an eigenray. This is the observation of~\cite{corichi2008} that we recalled in the introduction, now in its simplest form. The quantum condition imposes the analogue of the classical constraint, $c=0$, and in addition it requires the gauge vector field to vanish. Both requirements are imposed together, since by~\eqref{eq:fields} the squared residual is the squared mean plus $\hbar/2$ times the squared orbit speed. On semiclassical states the second term is of order $\hbar$, which is why it has no counterpart on $\Sigma$~\cite{corichi2008}. The physical state space is $\Pphy=\bigcap_a\PP(\ker\C_a)$. On each $\PP(\ker\C_a)$ the function $c_a$ vanishes to second order, whereas a classical constraint vanishes to first order on $\Sigma$. When the constraints generate a connected group $G$, $\dd c_a=0$ for all $a$ means the ray is fixed by $G$, and then $\Pphy=\mu^{-1}(0)\cap\PP(\Hil)^G$ with $\mu=(c_1,\dots,c_N)$~\cite{corichi2008}.

\paragraph{The physical projector.} For the projection arguments assume that $K_\hbar=\bigcap_a\ker\C_a\neq\{0\}$, and let $P_\hbar$ be its orthogonal projector. At $[\psi]\in\PP(K_\hbar)$ the tangent space is $K_\hbar\cap\psi^\perp$. If $v$ is a nonzero tangent vector, so is $Jv$, and $\Omega_\hbar(v,Jv)=g_\hbar(v,v)>0$. So the pullback of $\Omega_\hbar$ to $\Pphy$ is nondegenerate, and $\Pphy$ is a K\"ahler submanifold, in fact a totally geodesic one~\cite{corichi2008}. Since $\C_aP_\hbar=0$, the spectral theorem gives $P_\hbar\,\varphi(\C_a)=\varphi(0)P_\hbar$ for every Borel function $\varphi$. Hence every word $W$ in the real unitaries $e^{-it\C_a/\hbar}$ and, on their domains, the imaginary flows $e^{\tau\C_a/\hbar}$ satisfies
\begin{equation}\label{eq:PW}
\begin{gathered}
P_\hbar W=P_\hbar,\\
F_\hbar(\psi):=\frac{\|P_\hbar\psi\|^2}{\|\psi\|^2}=\sup_{\phi\in K_\hbar,\ \|\phi\|=1}\frac{|\langle\phi|\psi\rangle|^2}{\|\psi\|^2}=\cos^2\theta .
\end{gathered}
\end{equation}
Here $\theta$ is the Fubini--Study angle from $[\psi]$ to $\Pphy$, and the $g_\hbar$-distance is $\sqrt{2\hbar}\,\theta$; the invariance of the fibers of $[\psi]\mapsto[P_\hbar\psi]$ under the complexified gauge flows and the formula $F_\hbar=\cos^2\theta$ are in~\cite{corichi2008}. We call $F_\hbar$ the \emph{physical weight}. Its operational meaning is simple: if one tests whether a system prepared in the kinematical state $\psi$ is physical, by measuring the projector $P_\hbar$, the answer is yes with probability $F_\hbar(\psi)$. It is thus a probability defined by the kinematical inner product, not the norm of the normalized physical state. A bounded self-adjoint observable preserves the physical space if it maps $K_\hbar$ into itself. A probe that resolves a gauge coordinate need not have that property; its use below describes a kinematical preparation, not a measurement of a gauge coordinate by a physical observer.

\paragraph{The Gram matrix.} For several constraints the metric measures a matrix rather than a single speed:
\begin{equation}\label{eq:gram}
\begin{gathered}
\mathsf G_{ab}=g_\hbar(X_{c_a},X_{c_b})=\tfrac{2}{\hbar}\Re\langle d_a|d_b\rangle=\tfrac{2}{\hbar}\Re\Cov(\C_a,\C_b),\\
\Omega_\hbar(X_{c_a},X_{c_b})=\{c_a,c_b\}=\tfrac{1}{i\hbar}\langle[\C_a,\C_b]\rangle .
\end{gathered}
\end{equation}
For real coefficients $v^a$ one has $v^a\mathsf G_{ab}v^b=(2/\hbar)\|\sum_av^ad_a\|^2$, so the null space of $\mathsf G$ consists of the combinations with zero projective velocity at this ray, including those that act only by a phase. All entries vanish on $\Pphy$. Away from it the rank counts independent quantum velocities; identifying that rank with a classical leaf dimension requires control of which motions change the limiting probe readings (Sec.~\ref{sec:motion}). The bracket matrix in~\eqref{eq:gram} separates an uncertainty obstruction from mere packet dispersion: for a canonically normalized second-class pair $\Delta Q\,\Delta P\geq\hbar/2$, whereas for first-class directions the leading on-surface bracket vanishes.

\subsection{The physical weight along the gauge flows}\label{sec:weight}

We can now state the first result of this paper. It describes how the physical weight changes along the real and imaginary gauge flows, and it is the quantum counterpart of the classical relation~\eqref{eq:classKN}.

\begin{proposition}[exact identities]\label{prop:identities}
Let $\C_1,\dots,\C_N$ be self-adjoint with common kernel $K_\hbar\neq\{0\}$. At every ray with $F_\hbar>0$ and $\psi$ in the domains,
\begin{equation}\label{eq:identities}
X_{c_a}F_\hbar=0,\qquad \dd\log F_\hbar(\nabla c_a)=g_\hbar(\nabla c_a,\nabla\log F_\hbar)=-\tfrac{2}{\hbar}\,c_a .
\end{equation}
\end{proposition}
\begin{proof}
Since $\C_aP_\hbar=P_\hbar\C_a=0$, $\Cov(\C_a,P_\hbar)=-c_aF_\hbar$, a real number. Its imaginary part gives $\{c_a,F_\hbar\}=0$. Its real part gives $g_\hbar(\nabla c_a,\nabla F_\hbar)=(2/\hbar)\Re\Cov(\C_a,P_\hbar)=-(2/\hbar)c_aF_\hbar$.
\end{proof}

The real gauge flows preserve the weight, and the imaginary ones change it at a rate set by the constraint value. Several remarks are in order. First, \eqref{eq:identities} is the K\"ahler-geometric form of the operator statement $\C_aP_\hbar=0$. It contains no products of constraints, and its only input besides the $\C_a$ is $\Pphy$, through $F_\hbar=\cos^2\theta$. Second, it holds for any set of self-adjoint constraints with a common kernel. The constraints need not close, and they may have structure functions. This is the situation of general relativity, where group averaging is not available and where the master constraint programme was proposed for precisely this reason~\cite{thiemannMC}. Third, comparing with~\eqref{eq:classKN}, the identity says that $-\hbar\log F_\hbar$ plays in the quantum theory the role that the squared norm plays in the classical Kempf--Ness picture: its derivative along each imaginary gauge direction is twice the constraint value. For coherent states of a linear action the two functions agree exactly, up to a constant on each complexified orbit, as we shall see in Theorem~\ref{thm:linear}.

\begin{corollary}[one imaginary line at a time]\label{cor:line}
For every real combination $\C_\xi=\xi^a\C_a$, along $\tau\mapsto[e^{\tau\C_\xi/\hbar}\psi]$,
\begin{equation}\label{eq:line}
\frac{\dd}{\dd\tau}\log F_\hbar=-\frac{2}{\hbar}\langle\C_\xi\rangle_\tau,\qquad \frac{\dd^2}{\dd\tau^2}\log F_\hbar=-\frac{4}{\hbar^2}\Var_\tau(\C_\xi)\leq0 .
\end{equation}
\end{corollary}

So $\log F_\hbar$ is concave along each imaginary gauge line and is maximal where $\langle\C_\xi\rangle=0$. On $\{F_\hbar>0\}$ the quantum constraint surface $\mu^{-1}(0)$ is therefore the set of rays at which $F_\hbar$ is stationary along every imaginary gauge direction. This characterization uses no master function, and it is worth contrasting it with the one in~\cite{corichi2008}. There, the physical state space $\Pphy$ itself was characterized as the minimum set of the expectation value of the master constraint operator, a function that vanishes to second order on $\Pphy$. Here it is $\mu^{-1}(0)$, the quantum counterpart of $\Sigma$, that is characterized, by a first-order condition on the physical weight in which the constraints enter one at a time. Turning these one-parameter statements into a statement about a whole complexified orbit needs a group (Sec.~\ref{sec:projection}).

\paragraph{Imposing the constraints in sequence.} Let $P_a$ project onto $\ker\C_a$. By von Neumann's alternating-projection theorem~\cite{vonneumann}, $(P_1P_2\cdots P_N)^n\to P_\hbar$ strongly. The iterated retractions $[\psi]\mapsto[P_a\psi]$ thus converge to $[P_\hbar\psi]$ whenever $P_\hbar\psi\neq0$, at a rate set by the angles between the kernels. The constraints never need to be combined into one operator.

\section{Dirac's classification at finite \texorpdfstring{$\hbar$}{hbar}}\label{sec:dirac}

Classically, Dirac's classification of constraints into first and second class is a statement about the degeneracy of the symplectic form restricted to the constraint surface. On the quantum state space the expectation values $c_a$ are functions on a symplectic manifold, and one may ask the same question about their common zero set $\mu^{-1}(0)$, the finite-$\hbar$ counterpart of $\Sigma$. In this section we show that the answer reproduces Dirac's classification at finite $\hbar$, and we point out a feature with no classical counterpart: the answer depends on how the constraints are written, whereas $\Pphy$ and $F_\hbar$ do not. We end the section with a normal form near $\Pphy$, which shows where, in the space of rays, the classical gauge orbits should \emph{not} be sought.

\begin{proposition}[Dirac's classification on the state space]\label{prop:classification}
On the space of rays, \eqref{eq:gram} gives $\{c_a,c_b\}=(i\hbar)^{-1}\langle[\C_a,\C_b]\rangle$. Hence:
\begin{enumerate}[label=(\arabic*)]
\item \emph{Structure constants}, $[\C_a,\C_b]=i\hbar f_{ab}{}^c\C_c$. Then $\{c_a,c_b\}=f_{ab}{}^cc_c$ vanishes on $\mu^{-1}(0)$. Where $\dd\mu$ has rank $N$, which is where the stabilizer is discrete, $\mu^{-1}(0)$ is coisotropic of codimension $N$, and its characteristic distribution is spanned by the $X_{c_a}$, the tangents of the quantum gauge orbits.
\item \emph{Central anomaly}, $[\C_a,\C_b]=i\hbar(f_{ab}{}^c\C_c+\hbar A_{ab})$. Then $\{c_a,c_b\}=\hbar A_{ab}$ on $\mu^{-1}(0)$, which is not coisotropic where $A\neq0$; it behaves as a second-class surface with Dirac matrix $\hbar A$. Moreover $K_\hbar=\{0\}$, because $\langle\psi|[\C_a,\C_b]|\psi\rangle=0$ on a common kernel.
\item \emph{Operator structure functions}, $[\C_a,\C_b]=i\hbar\,\tfrac12\{\hat f_{ab}{}^c,\C_c\}$. Then on $\mu^{-1}(0)$ the bracket is $\Re\Cov(\hat f_{ab}{}^c,\C_c)$, which is generally nonzero: the surface fails to be coisotropic even without an anomaly.
\end{enumerate}
\end{proposition}

Case (1) is the classical first-class geometry, realized by expectation values. Case (2) is the model of~\cite{corichiAnom}: a central term, which for the generators of a symmetry would only make the representation projective, turns the quantum constraint surface into a second-class one and leaves no physical states. Case (3) is the situation of gravity, and it deserves a comment. In general relativity the constraint algebra closes with structure functions, and Proposition~\ref{prop:classification} says that one should not expect the quantum constraint surface to be coisotropic at finite $\hbar$, even in the absence of anomalies. First-class character can then at most emerge in the limit $\hbar\to0$, on suitable families of states, which is the point of view of effective constraints~\cite{bojowald}. The situation already occurs in a simple mechanical model, which also shows that the obstruction depends in part on the chosen presentation of the constraints.

\paragraph{Presentation dependence.} On $T^*T^3$ with angles $x,y,u$ take (the model of Sec.~\ref{sec:structure})
\begin{equation}\label{eq:T3}
\begin{gathered}
\C_1=\hat p_x+g'(x)\hat p_u,\qquad \C_2=f(x)\hat p_y,\qquad g=\sin x,\quad f=2+\sin x,\\
\frac{[\C_1,\C_2]}{i\hbar}=-f'(x)\hat p_y=-\frac{f'}{f}\,\C_2 .
\end{gathered}
\end{equation}
The structure function $f'/f$ commutes with $\C_2$, so on $c_2=0$ the bracket is $\{c_1,c_2\}=-\Cov(f'/f,\C_2)$. Take $\psi=a(x)e^{iy}+b(x)e^{-iy}$ with real $a,b$, $a^2-b^2=h(x)$ and $a^2+b^2=\rho$, a constant with $\rho\geq\max|h|$. Then $c_1=0$ identically, and
\begin{equation}\label{eq:T3values}
c_2=\frac{\hbar}{\rho}\langle fh\rangle_x,\qquad \langle\hat p_y\rangle=\frac{\hbar}{\rho}\langle h\rangle_x,\qquad \{c_1,c_2\}=-\frac{\hbar}{\rho}\langle f'h\rangle_x,
\end{equation}
where $\langle\cdot\rangle_x$ is the average over $x$. Two choices make the point.
\begin{itemize}
\item For $h=\cos x$, $c_1=c_2=\langle\hat p_y\rangle=0$, but $\{c_1,c_2\}=-\hbar/(2\rho)\neq0$. In the commuting frame $(\C_1,\hat p_y)$ the bracket vanishes identically. The same ray lies on both zero sets, which are coisotropic there in one frame and not in the other.
\item For $h=1-4\sin x$, $c_2=0$ but $\langle\hat p_y\rangle=\hbar/\rho\neq0$, so the two frames have different zero sets.
\end{itemize}
Yet $f>0$, so $\ker\C_2=\ker\hat p_y$, and the common kernel, $P_\hbar$, $\Pphy$ and $F_\hbar$ are identical in the two frames. The surface $\mu^{-1}(0)$ is a property of the chosen presentation, while $F_\hbar$ is a property of the constraints as a system; this is why the projection reading of Sec.~\ref{sec:projection} is built on $F_\hbar$. On localized packets the obstruction is small: by Cauchy--Schwarz, $|\Cov(f'/f,\C_2)|\leq\Delta(f'/f)\,\Delta\C_2=O(\hbar)$ for packets of the type used in Sec.~\ref{sec:examples}. This is the finite-$\hbar$ test of first-classness that Sec.~\ref{sec:motion} requires: the bracket on $\mu^{-1}(0)$ must vanish as $\hbar\to0$ on the packet family.

\begin{proposition}[normal form at $\Pphy$]\label{prop:normal}
Let $[\Psi_0]\in\Pphy$. Parametrize the affine chart by $\delta\perp\Psi_0$, $\delta\mapsto[\Psi_0+\delta]$, and split $\delta=\delta_K+\delta_\perp$ with $\delta_K\in K_\hbar$ and $\delta_\perp\in K_\hbar^\perp$. Then, exactly on the whole chart,
\begin{equation}\label{eq:normal}
c_a([\Psi_0+\delta])=\frac{\langle\delta_\perp|\C_a|\delta_\perp\rangle}{1+\|\delta\|^2},\qquad F_\hbar([\Psi_0+\delta])=\frac{1+\|\delta_K\|^2}{1+\|\delta\|^2}.
\end{equation}
\end{proposition}
\begin{proof}
The cross terms vanish because $\C_a\Psi_0=0$ and $\C_a\delta_K=0$.
\end{proof}

So $\dd\mu$ vanishes on $\Pphy$, which is the deepest singular stratum of $\mu^{-1}(0)$, in agreement with the description of $\Pphy$ in~\cite{corichi2008} as the deepest stratum of the K\"ahler quotient of $\PP(\Hil)$. Near it $\mu$ is the quadratic moment map of the constraints acting on the unphysical states $K_\hbar^\perp$, and the quantum gauge orbits are rotations there. The classical gauge orbits, which have finite size, are not near $\Pphy$. They are carried by packets far from it, on which $F_\hbar$ is of order $\hbar^{k/2}$ (Sec.~\ref{sec:landscape}). This is the first indication that the gauge orbits are to be found in the relation between the kinematical and the physical states, and not in a neighborhood of $\Pphy$.

\section{Gauge directions from motion}\label{sec:motion}

Let us now turn to the first of our two readings. The question it addresses is operational. Suppose we are given kinematical preparations that are localized in some classical phase space. How can we tell, from the response of kinematical probes, that the motion induced by a constraint unitary is a gauge motion, that is, a motion along the null directions of the classical constraint surface? The answer, stated in Proposition~\ref{prop:local}, is that under explicit correspondence conditions the velocities that constraint unitaries impart to packet labels span the characteristic distribution of $\Sigma$. We then spell out what the hypotheses carry and explain why the quantum theory needs no second quotient. Equal projection of the translated packets is consistent with all this, but, as we shall see at the end of the section, it does not single out gauge-related points.

\paragraph{Setting.} Suppose an operational choice has been made of a smooth finite-dimensional label space $\Gamma$, rays $s_\hbar(z)$ and kinematical probes $\hat B^\mu$. Denote the probe symbols by $b^\mu(z)=\lim\langle\hat B^\mu\rangle_{s_\hbar(z)}$. On a patch assume that their differentials span the cotangent space, so that readings give local coordinates, possibly with redundancies. The constraint response measured at $z$ is
\begin{equation}\label{eq:response}
R_a^\mu(z)=\lim_{\hbar\to0}\ \big\langle[\hat B^\mu,\C_a]/(i\hbar)\big\rangle_{s_\hbar(z)} .
\end{equation}
Choose $d=\dim\Gamma$ probe symbols with independent differentials and use their values as coordinates. In those coordinates the $d$ numbers $R_a^\mu$ are the components of a candidate velocity $Y_a$; in any other chart one solves $(\partial_ib^\mu)Y_a^i=R_a^\mu$. Further probes test this assignment: their responses must equal $\dd b(Y_a)$. Thus one can infer a direction before testing whether it is the Hamiltonian direction of a classical constraint. A separating set of values distinguishes points; the differential rank condition distinguishes tangent vectors.

\begin{proposition}[local identification]\label{prop:local}
Suppose $s_\hbar^*\Omega_\hbar\to\omega$ locally uniformly with smooth symplectic limit, and $c_a\circ s_\hbar\to C_a$ in $C^1$ on compact sets, with smooth limits. Suppose the response identity $R_a^\mu=\{b^\mu,C_a\}_\omega$ holds for the spanning probes. Then the observed response has the unique tangent representative $X_{C_a}$. Suppose further that the common zero set $\Sigma$ is regular, with independent defining differentials, and that every bracket $\{C_a,C_b\}$ vanishes on $\Sigma$. Then on $\Sigma$ the span of the responses is precisely $\ker(\omega|_{T\Sigma})$.
\end{proposition}
\begin{proof}
On a compact patch, inversion of the pulled-back forms is continuous for small $\hbar$, so the Hamiltonian fields of the pulled-back functions tend to $X_{C_a}$. By the response identity this field solves $\dd b^\mu(Y)=R_a^\mu$ for all $\mu$, and full differential rank gives uniqueness. The conormal at $z\in\Sigma$ is spanned by the $\dd C_a(z)$, so $(T_z\Sigma)^\omega$ is spanned by the $X_{C_a}(z)$. First-classness makes every such field tangent. Their span is the null space of the pullback, has constant rank, and integrates to the characteristic foliation.
\end{proof}

\paragraph{What the hypotheses carry.} The response identity is a test, not a consequence of pulling back $\Omega_\hbar$: a quantum trajectory can leave the packet family immediately. Two strengthenings make the hypotheses checkable.
\begin{itemize}
\item \emph{K\"ahler calibration.} We ask, in addition, that $s_\hbar^*g_\hbar\to g$ with $g=\omega(\cdot,J\cdot)$ for a compatible complex structure $J$ on $\Gamma$. The circle packets of Sec.~\ref{sec:torus} satisfy this with $g=S\,\dd\theta^2+\dd p^2/S$, where $S$ is the action scale. Calibration is what fixes the power of $\hbar$ in the weight (Sec.~\ref{sec:volume}).
\item \emph{Toeplitz probes.} On a K\"ahler-calibrated family the natural probes are Berezin--Toeplitz operators $T_b$. In the sign convention of~\eqref{eq:gram} their commutators satisfy $[T_b,T_c]=i\hbar\,T_{\{b,c\}}+O(\hbar^2)$~\cite{bms}. We expect that the response identity then follows whenever each $\C_a$ agrees with $T_{C_a}$ up to $O(\hbar)$ on the family. Proving this, with the domain conditions, is a first task of the programme.
\end{itemize}
To establish a finite orbit one also needs a propagation statement, for example
\begin{equation}\label{eq:propagation}
\big\langle U_a(t)^*\hat B^\mu U_a(t)\big\rangle_{s_\hbar(z)}\longrightarrow b^\mu(\Phi_a^tz)
\end{equation}
on bounded time intervals, where $\Phi_a^t$ is the proposed classical flow. To use Theorem~\ref{thm:resolved}, test second moments as well, or establish localization in norm. For a word in the unitaries, propagation must hold along that word; iterating a result about unperturbed initial packets without further control would be insufficient. First-classness should likewise be tested on kinematical states. Proposition~\ref{prop:classification} gives the test at finite $\hbar$, and Sec.~\ref{sec:dirac} shows that it is presentation dependent.

\paragraph{Why there is no second quantum quotient.} For any constraint word $W$, \eqref{eq:PW} gives $P_\hbar W\psi=P_\hbar\psi$. When it exists, even the normalized projected ray is the same for all translated preparations. Their limiting kinematical readings can nevertheless give distinct labels $z$ and $\Phi_a^tz$ on $\Sigma$, and Proposition~\ref{prop:local} says precisely when their separation is along a null direction there. Nondegeneracy of the physical quantum two-form and degeneracy of the classical restricted form are statements about different spaces.

In the language of~\cite{abc2005}, the packets $s_\hbar(z)$ are kinematical semiclassical states and their projections are physical semiclassical states. The result of this section can then be phrased as follows. The information lost in the projection, the location of the packet along its gauge orbit, is precisely the information that the kinematical probes read, and Proposition~\ref{prop:local} states the conditions under which that location is a point of a classical gauge orbit.

\begin{remark}[equal projection is not a gauge criterion]\label{rem:equal}
The words in~\eqref{eq:PW} include imaginary flows. A packet pushed along $e^{\tau\C_a/\hbar}$ also has the same projection, but its label moves off $\Sigma$, as the circle example of Sec.~\ref{sec:intro} shows. Equal projection therefore identifies points of complexified gauge orbits, a larger set than the gauge leaves. Section~\ref{sec:projection} shows how $F_\hbar$ picks out $\Sigma$ inside each such set, and with it the real gauge orbits.
\end{remark}

\section{The constraint surface and the imaginary gauge directions from projection}\label{sec:projection}

We now come to the second reading, which is the core of this paper. In the motion reading the gauge orbits are traced by kinematical probes. Here we ask instead what can be learned from the physical projector alone, together with a family of kinematical states on which the gauge group acts. The classical model is the holomorphic description of reduction recalled in Sec.~\ref{sec:classical}, and the result is its faithful quantum counterpart. For a compact gauge group, each complexified gauge orbit lies in one projection fiber and contains exactly one real gauge orbit of maximal weight, the one on which all constraint expectations vanish. With a packet family on which the complexified group acts, this recovers $\Sigma$, its gauge orbits and the reduced space exactly, at every $\hbar$. This section has three parts: a general theorem for compact groups, its application to linear actions with coherent states, and its application to heat-kernel packets on cotangent bundles of compact groups.

\paragraph{Setting.} Let $G$ be compact and connected, with Lie algebra $\g$. Suppose the $\C_\xi=\xi^a\C_a$ generate a unitary representation $U$ of $G$, with $U(\exp\theta\xi)=e^{-i\theta\C_\xi/\hbar}$. Since $G$ is connected, $K_\hbar$ is the space of invariant vectors. Call $\Psi$ \emph{entire} if its isotypic components satisfy $\sum_\lambda e^{R|\lambda|}\|\Psi_\lambda\|<\infty$ for every $R$. On entire vectors $U$ extends holomorphically to the complexified group $\GC$, with $U(\exp i\tau\xi)=e^{\tau\C_\xi/\hbar}$; $G$-finite vectors, coherent states and the heat-kernel packets below are entire. The map $\mu=(c_a)$ is the moment map of the action of $G$ on $\PP(\Hil)$. The complexified orbit of $[\Psi]$ is $\mathcal O=\{[U(h)\Psi]:h\in\GC\}$, and its tangent space is spanned by the real and imaginary gauge directions.

Since $P_\hbar$ is invariant, $\|P_\hbar U(h)\Psi\|$ does not depend on $h$, and on a complexified orbit the physical weight is a constant multiple of $1/\|U(h)\Psi\|^2$. The logarithm of $\|U(h)\Psi\|^2$ is the Kempf--Ness function of $\Psi$~\cite{kempfness,woodward}. The following theorem is therefore the Kempf--Ness theorem, extended to entire vectors and read through the physical weight; item~(1) and the Gaussian law in item~(5) are what this reading adds.

\begin{theorem}[Kempf--Ness theorem for the physical weight]\label{thm:KN}
Let $\Psi$ be entire with $P_\hbar\Psi\neq0$.
\begin{enumerate}[label=(\arabic*)]
\item $P_\hbar U(h)\Psi=P_\hbar\Psi$ for all $h\in\GC$. Hence $\mathcal O$ lies in one fiber of the retraction $[\psi]\mapsto[P_\hbar\psi]$, and $F_\hbar([U(h)\Psi])=\|P_\hbar\Psi\|^2/\|U(h)\Psi\|^2$ on $\mathcal O$.
\item $\Phi(h)=\log\|U(h)\Psi\|^2$ is invariant under $h\mapsto kh$ for $k\in G$. Along every curve $\tau\mapsto\exp(i\tau\xi)h$ it satisfies~\eqref{eq:convex}; in particular it is convex.
\item A ray of $\mathcal O$ is a critical point of $F_\hbar$ on $\mathcal O$ if and only if $\mu=0$ there, if and only if it is a maximum of $F_\hbar$ on $\mathcal O$. The set $\mathcal O\cap\mu^{-1}(0)$ is empty or a single $G$-orbit.
\item For $G$-finite $\Psi$, $\mathcal O\cap\mu^{-1}(0)$ is nonempty if and only if the orbit $\GC\cdot\Psi$ is closed. For a torus this holds if and only if $0$ lies in the relative interior of the convex hull of the weights of $\Psi$ ($0$ is a weight because $P_\hbar\Psi\neq0$). Otherwise the supremum of $F_\hbar$ on $\mathcal O$ is approached along a one-parameter subgroup, towards a ray with fewer weights.
\item Near a maximum, along $\exp(i\xi)$, the constraint values are $c=(2/\hbar)\Cov\,\xi+O(|\xi|^2)$, and the Gaussian law~\eqref{eq:gauss} holds, where $\Cov$ is the real covariance matrix of the $\C_a$ at the maximum, on a complement of the stabilizer algebra.
\end{enumerate}
\begin{gather}
\frac{\dd\Phi}{\dd\tau}=\frac{2}{\hbar}\langle\C_\xi\rangle,\qquad \frac{\dd^2\Phi}{\dd\tau^2}=\frac{4}{\hbar^2}\Var(\C_\xi)\quad\text{in the state }U(\exp(i\tau\xi)h)\Psi,\label{eq:convex}\\
\log F_\hbar=\log F_\hbar^{\max}-\tfrac12\,c^{\mathsf T}\Cov^{-1}c+O(|c|^3).\label{eq:gauss}
\end{gather}
\end{theorem}
\begin{proof}
(1) $P_\hbar$ commutes with $U(k)$, and $P_\hbar e^{\C_\xi/\hbar}=P_\hbar$ by the spectral theorem; every $h$ is $k\exp(i\xi)$ by the Cartan decomposition. (2) $U(k)$ is unitary. Along the curve, $\|e^{\tau\C_\xi/\hbar}\phi\|^2=\int e^{2\tau\lambda/\hbar}\,\dd\langle\phi|E_\lambda\phi\rangle$ is the Laplace transform of a positive measure, so its logarithm is convex by H\"older's inequality. (3) The function $\Phi$ descends to the symmetric space $G\backslash\GC$, whose geodesics are the curves in (2). Being geodesically convex, its critical points are its minima. Convexity is strict except along $\xi$ with $\Var(\C_\xi)=0$; at a critical point that means $\C_\xi\phi=0$, so $\xi$ lies in the stabilizer algebra, and the minima form one $G$-orbit of rays. The first derivative in~\eqref{eq:convex} vanishes for all $\xi$ exactly when $\mu=0$. (4) This is the Kempf--Ness theorem~\cite{kempfness} in the finite-dimensional space spanned by the isotypic components of $\Psi$, together with the Hilbert--Mumford criterion~\cite{git}. (5) Expand~\eqref{eq:convex} to second order.
\end{proof}

Let us pause to discuss the meaning of Theorem~\ref{thm:KN}. In~\cite{corichi2008} we contrasted two ways of selecting states in $\PP(\Hil)$ by means of the constraints. The first mimics the classical procedure: impose $\mu=0$ and take the quotient by $G$, which gives the K\"ahler quotient of the space of rays. The second is Dirac's: project onto the invariant subspace. We showed there that the two are very different, since $\Pphy$ is only the deepest stratum of the K\"ahler quotient. Theorem~\ref{thm:KN} shows that they are nevertheless closely related within each complexified orbit. Dirac projection collapses the whole complexified orbit $\mathcal O$ to a single physical ray, while the K\"ahler quotient picks from $\mathcal O$ a single gauge orbit, the one on which $\mu=0$. The rays the K\"ahler quotient picks are precisely those that Dirac projection loses least. In other words, among all kinematical states with the same physical image, those on $\mu^{-1}(0)$ have the largest probability of passing the test ``is it physical?''.

Item~(5) makes this quantitative. Away from its maximum, the physical weight decreases as a Gaussian in the constraint values, measured in units of their quantum spread. For semiclassical states, whose constraint variances are of order $\hbar$, the quantum constraint surface is thus a fuzzy version of $\Sigma$, of width of order $\sqrt\hbar$ in the constraint values, and $F_\hbar$ is the profile across it. Finally, item~(4) says what happens when a complexified orbit does not meet $\mu^{-1}(0)$: the weight then has no maximum on the orbit, and its supremum is approached as the state degenerates. We shall meet an example in Sec.~\ref{sec:linear}.

The general theorem refers to the space of rays. To recover the classical gauge geometry we need, in addition, a family of kinematical states labeled by points of the classical phase space and covariant under the complexified group. We now consider the two cases in which such a family is available in closed form.

The next theorem assembles classical ingredients: the Bargmann reproducing kernel, coherent-state matrix elements of the physical projector as in Klauder's approach~\cite{klauder1997}, the separation of closed orbits by invariants~\cite{git}, the Kempf--Ness theorem and the K\"ahler potential of a quotient~\cite{hklr}. Its content is the converse direction: $\Sigma$, its gauge orbits and the reduced K\"ahler structure are determined by $\Pphy$ and the coherent family alone.

\begin{theorem}[linear actions: recovery from $\Pphy$]\label{thm:linear}
Let $\Gamma=\Cx^n$ with $\omega=i\sum\dd z\wedge\dd\bar z$. Unnormalized coherent states are $\cs z=\exp(z\cdot\hat a^\dagger/\sqrt\hbar)\ket0$, with $(w\cs z=e^{\bar w\cdot z/\hbar}$, and $\ket z=e^{-|z|^2/2\hbar}\cs z$. The map $z\mapsto[\ket z]$ pulls the Fubini--Study structure back to $\omega$ and $g=2|\dd z|^2$ exactly. A compact $G$ acts on $\Cx^n$ by a unitary representation $u$, and $U$ is its second quantization, so $U(h)\cs z=\cs{u(h)z}$ for $h\in\GC$. The constraints are the normal-ordered quadratics $\C_\xi=i\hbar\,\hat a^\dagger\cdot\dd u(\xi)\hat a$, whose coherent expectations are exactly the classical moment map. Write $F_\hbar(z)=\langle z|P|z\rangle$ for the physical weight of the coherent state and $\pi(z)=[P\cs z]$. Then, for every $\hbar>0$:
\begin{enumerate}[label=(\arabic*)]
\item $(z|P\cs z=\int_Ge^{\bar z\cdot u(g)z/\hbar}\,\dd g$, and $P\cs{u(h)z}=P\cs z$ for $h\in\GC$. So $F_\hbar(u(h)z)=(z|P\cs z\,e^{-|u(h)z|^2/\hbar}$.
\item $\pi(z)=\pi(z')$ if and only if the closures of $\GC\cdot z$ and $\GC\cdot z'$ meet. On the polystable set $\Gamma^{\mathrm{ps}}$, the points with closed complexified orbit, the fibers of $\pi$ are exactly the complexified gauge orbits.
\item On each fiber, $F_\hbar$ is maximal exactly on the gauge orbit $G\cdot z\cap\mu^{-1}(0)$. Therefore $\Sigma=\{z:F_\hbar(z)\text{ is maximal on the fiber of }\pi\text{ through }z\}$, for every $\hbar$; points of $\Sigma$ are polystable.
\item The classical gauge orbits are the fibers of $\pi$ intersected with $\Sigma$.
\item $[z]\mapsto\pi(z)$ maps $\Sigma/G$ injectively into $\Pphy$. The Fubini--Study form of $\Pphy$ pulls back to $i\hbar\partial\bar\partial\log(z|P\cs z$, so $\mathcal K_\hbar=\hbar\log(z|P\cs z=|z|^2+\hbar\log F_\hbar$ is an exact physical K\"ahler potential. Its leading term $|z|^2$, restricted to $\Sigma$, is a K\"ahler potential for $\omega_\red$, as for any K\"ahler quotient at level zero~\cite{hklr}. At a point of $\Sigma$ with stabilizer $G_z$, Laplace's method gives~\eqref{eq:laplace}.
\item Along every imaginary gauge curve, $\dd\log F_\hbar/\dd\tau\,(\exp(i\tau\xi)z)=-(2/\hbar)\,\mu_\xi(\exp(i\tau\xi)z)$. So the constraint functions are recovered from $F_\hbar$ off $\Sigma$ as well.
\end{enumerate}
\begin{equation}\label{eq:laplace}
F_\hbar(z)=\frac{\vol(G_z)}{\vol(G)}\,\frac{(2\pi\hbar)^{k/2}}{\sqrt{\det M_\perp(z)}}\,\big(1+O(\hbar)\big),\qquad M_{ab}(z)=\Re\langle\xi_az,\xi_bz\rangle .
\end{equation}
Here $k=\dim G-\dim G_z$, the $\xi_a$ run over an orthonormal basis of a complement of the stabilizer algebra, and volumes are taken in the same metric. For a free orbit, $G_z$ is trivial and $k=\dim G$.
\end{theorem}
\begin{proof}
(1) $P=\int_GU(g)\,\dd g$ and $(w|U(g)\cs z=e^{\bar w\cdot u(g)z/\hbar}$. Each degree-$m$ component of $\cs z$ lies in the finite-dimensional invariant space $S^m(\Cx^n)$, so Theorem~\ref{thm:KN}(1) applies degree by degree. (2) The degree-$m$ part of $P\cs z$ is, up to a constant, the invariant projection of $z^{\otimes m}$. Its pairing with an invariant polynomial $q$ of degree $m$ is proportional to $q(z)$, and the pairing between invariants is perfect for compact $G$. If $\pi(z)=\pi(z')$, then $P\cs z=\lambda P\cs{z'}$; the vacuum components give $\lambda=1$, so $q(z)=q(z')$ for every invariant $q$. Invariants of a reductive group separate closed orbits~\cite{git}. (3) Item~(1) and the classical Kempf--Ness theorem for the norm $|u(h)z|^2$. (4) Items~(2) and~(3). (5) $z\mapsto P\cs z$ is holomorphic, so the pulled-back form is $i\hbar\partial\bar\partial\log\|P\cs z\|^2$. On $\Sigma$, $\bar z\cdot u(\exp\xi)z=|z|^2-\tfrac12|\dd u(\xi)z|^2+O(|\xi|^3)$, and the maximum of $\Re\,\bar z\cdot u(g)z$ is attained only on $G_z$. (6) Theorem~\ref{thm:KN}(2).
\end{proof}

\paragraph{Input and output.} The input is the K\"ahler space of rays, its subset $\Pphy$, and the coherent family, a K\"ahler embedding $\Gamma\to\PP(\Hil)$ covariant under $\GC$. The output is $\Sigma$, its null foliation, the complexified orbits, and $\Gamma_\red$ with its symplectic form. Neither the constraint operators nor a master function appear; the gauge group enters only through the covariance of the family. Item~(5) also identifies the first quantum correction to the reduced K\"ahler potential: on $\Sigma$, $\hbar\log F_\hbar$ is minus $\hbar$ times the logarithm of the gauge-orbit volume, up to a constant (Proposition~\ref{prop:volume}). The orbit volume appears in the same way in the diagonal of invariant Szeg\H{o} and Bergman kernels on the zero level set~\cite{paoletti2005,paoletti2008,mazhang}, and it is the obstruction to asymptotic unitarity of the Guillemin--Sternberg map that the metaplectic correction removes~\cite{hallkirwin}.

Two further features of Theorem~\ref{thm:linear} are worth stressing. First, it holds at every $\hbar$, not only asymptotically. The reason is that, for linear actions, the coherent-state expectations of the normal-ordered constraints are exactly the classical moment map, and the complexified group acts on coherent states exactly as it acts on their labels. By item~(1), $-\hbar\log F_\hbar(u(h)z)=|u(h)z|^2-\hbar\log(z|P\cs z$ along each complexified orbit, so the quantum Kempf--Ness function of Theorem~\ref{thm:KN} is the classical one of Sec.~\ref{sec:classical}, up to a constant. Second, the theorem may be viewed as ``quantization commutes with reduction''~\cite{guilleminsternberg} as seen from the space of states. The invariant subspace of Fock space is a holomorphic quantization of $\Gamma_\red$, and the projected coherent states $P\cs z$, which depend only on the complexified orbit of $z$, are its reproducing-kernel vectors. What the theorem adds is that the physical weight singles out, within each such orbit, the representatives that lie on $\Sigma$.

\begin{theorem}[heat-kernel packets on $T^*G$]\label{thm:hall}
Let $H_t$ be the central heat kernel of $\sum_aY_a^2$ at the identity, with $\int H_t=1$ and convention $e^{t\sum Y_a^2}$, continued analytically to $\GC$. Identify $T^*G\cong\GC$ by $(g,Y)\mapsto g\,e^{iY}$, as in Hall's coherent-state transform~\cite{hall1994,hall2018}. For $z\in\GC$ put $\psi_z(g)=H_t(z^{-1}g)$. Let the constraints be $\C_a=-i\hbar Y_a$ for the left-invariant fields $Y_a$, which generate right multiplication, so $(U(h)f)(g)=f(gh^{-1})$. Then:
\begin{enumerate}[label=(\arabic*)]
\item $U(h)\psi_z=\psi_{zh}$ for all $h\in\GC$, and $P_\hbar\psi_z=1$, the constant function.
\item With $z=g\,e^{iY}$,
\begin{equation}\label{eq:hallnorm}
\begin{gathered}
\|\psi_{g e^{iY}}\|^2=H_{2t}\big(e^{-2iY}\big)=\sum_\lambda d_\lambda\,e^{-2tc_\lambda}\,\chi_\lambda\big(e^{2iY}\big),\\
\text{SU(2): }\ \chi_j\big(e^{2iY}\big)=\frac{\sinh((2j+1)|y|)}{\sinh|y|},
\end{gathered}
\end{equation}
where $c_\lambda$ is the Casimir eigenvalue ($j(j+1)$ for SU(2)) and $Y=\sum y_a\tau_a$ with $\tau_a=-\tfrac i2\sigma_a$.
\item $F_\hbar=1/\|\psi_z\|^2$ is maximal exactly on the zero section $Y=0$, a single $G$-orbit of labels; $\GC$ acts transitively, so there is one complexified orbit. The zero section of $T^*G$ is the constraint surface, and the packets of maximal weight are the real-centered heat-kernel packets of Sec.~\ref{sec:su2}.
\item $\dd\log F_\hbar/\dd\tau=-(2/\hbar)\langle\C_\xi\rangle$ along $z\,e^{i\tau\xi}$, and the Gaussian law~\eqref{eq:gauss} holds near $Y=0$.
\end{enumerate}
\end{theorem}
\begin{proof}
$H_t$ is a class function, so $H_t(z^{-1}gh^{-1})=H_t((zh)^{-1}g)$; this gives (1) after analytic continuation, and $\int\psi_z=1$ continues from real $z$. For (2), the conjugation of $\GC$ fixing $G$ gives $\overline{\psi_z}=\psi_{\bar z}$, and the semigroup property gives $\|\psi_z\|^2=H_{2t}(z^{-1}\bar z)$ with $z^{-1}\bar z=e^{-2iY}$, as for Hall's coherent states~\cite{hall1994}. For (3), each character $\chi_\lambda(e^{2iY})$ is a sum of $e^{2w(Y)}$ over the weights $w$ of $\lambda$. The weights are symmetric under $w\mapsto-w$, so the sum is a sum of $\cosh$ terms; it is at least $d_\lambda$, with equality for all $\lambda$ only at $Y=0$. (4) is Theorem~\ref{thm:KN}(2) and (5).
\end{proof}

With a spectator factor, $\psi_z\otimes\phi_w$ with $\phi_w$ any coherent family on $\Gamma_\red$, the fibers of projection are $\GC\times\{w\}$, $\Sigma$ is the zero section times $\Gamma_\red$, and the physical K\"ahler potential is that of $\Gamma_\red$. For $G=U(1)^r$, $\psi_z$ is the periodic Gaussian of Sec.~\ref{sec:torus} with heat time $\varepsilon/2$; its momentum label is the imaginary part of the complexified angle, with $|Y|=|p|/S$. The sphere packets of Sec.~\ref{sec:sphere} are not of this form, since $S^2$ is not a group; their Kempf--Ness analysis needs coherent states on $T^*S^2$ (Sec.~\ref{sec:discussion}).

For $G=SU(2)$ the packets of Theorem~\ref{thm:hall} are the coherent states used in loop quantum gravity on a single edge, where they are known as heat-kernel or complexifier coherent states~\cite{thiemannGCS,thiemannwinkler}. Theorem~\ref{thm:hall} describes them from the point of view of a gauge constraint acting by right translations. On a graph, the Gauss constraint at each vertex acts by left or right translations, according to orientation, on the edges that meet there, and the SU(2) closure constraint of open problem~\ref{op:closure} in Sec.~\ref{sec:discussion} is the Gauss constraint at a single vertex.

\section{The physical weight landscape}\label{sec:landscape}

The results of the two previous sections can be summarized by saying that the physical weight is constant along the real gauge orbits and peaks on $\Sigma$, where it is Gaussian in the constraint values. In this section we study the values it takes, and this is where the two readings meet. We first show that a kinematical state that resolves points of a gauge orbit, in the sense of the motion reading, must have a small physical weight. We then show that, for compact groups, the phases of the overlaps between translates carry the constraint values, and that the weight on $\Sigma$ is one over the number of Planck cells that cover the gauge orbit. K\"ahler-calibrated packets therefore attain the bound.

\subsection{Resolved translates}

Fix a normalized packet $\psi_\hbar$. Choose any finite list of constraint words $W_j$ and write $\psi_j=W_j\psi_\hbar$. The list need not arise from a Lie group with fixed structure constants. Let $O_{ij}=\langle\psi_i|\psi_j\rangle$ be their overlap matrix; relative phases refer to the actual unitary words, not to independently chosen ray representatives.

\begin{theorem}[resolved translates]\label{thm:resolved}
For each $N$ and all such words,
\begin{equation}\label{eq:resolved}
F_\hbar(\psi_\hbar)\leq\frac1{N^2}\sum_{i,j}O_{ij}\leq\frac1N+\frac1{N^2}\sum_{i\neq j}|O_{ij}| .
\end{equation}
Suppose that for every fixed $N$ there are $N$ translates for which each pair can be distinguished by a bounded self-adjoint probe, possibly pair dependent: its limiting means in the two translates differ and its two variances tend to zero. Then $F_\hbar(\psi_\hbar)\to0$.
\end{theorem}
\begin{proof}
All translates have the same projection, so their average $\bar\psi=N^{-1}\sum_j\psi_j$ has $P_\hbar\bar\psi=P_\hbar\psi_\hbar$. Taking the squared norm gives the first inequality; the sum of overlaps is real and nonnegative. The second follows by taking absolute values. For two normalized vectors with probe means $b_i,b_j$ and standard deviations $\sigma_i,\sigma_j$, insert the probe on either side of the matrix element and use self-adjointness:
\begin{equation}\label{eq:probe}
|b_i-b_j|\,|\langle\psi_i|\psi_j\rangle|\leq\sigma_i+\sigma_j .
\end{equation}
Each off-diagonal term therefore tends to zero for fixed $N$. Taking $\limsup_{\hbar\to0}$ in~\eqref{eq:resolved} first gives $1/N$; only then let $N$ grow.
\end{proof}

Theorem~\ref{thm:resolved} has a simple physical interpretation. A kinematical state cannot be both sharply localized along a gauge orbit and physical: if $N$ points of the orbit can be told apart, the probability that the state passes the test ``is it physical?'' is at most of order $1/N$. This is a kind of complementarity between gauge localization and physicality, and it holds for any constraint words, with no group structure and no assumption on the constraint algebra. It also explains why the classical gauge orbits could not be found near $\Pphy$ in Sec.~\ref{sec:dirac}: the states that carry them are, necessarily, far from it.

Let us note a few technical points. The order of the two limits matters. One need not resolve a growing number of points uniformly, nor bound the rate at which variances vanish; an open arc of one orbit suffices if each fixed finite subset can be resolved. Different mean readings without sharpness do not imply orthogonality. Nor does small $F_\hbar$ by itself establish gauge motion: a fixed ray in a nonzero eigenspace has $F_\hbar=0$.

\begin{corollary}[a quantitative orbit patch]\label{cor:patch}
Let $D\subset\R^r$ be a bounded parameter patch of positive volume, with strongly measurable translates $\psi_s=W(s)\psi_\hbar$. In dimensionless parameter coordinates suppose, uniformly, that $|\langle\psi_s|\psi_t\rangle|\leq A\,e^{-a|s-t|^2/\varepsilon}+R_\varepsilon$ with $a,A>0$ and $R_\varepsilon\geq0$. Then $F_\hbar\leq A|D|^{-1}(\pi\varepsilon/a)^{r/2}+R_\varepsilon$.
\end{corollary}
\begin{proof}
Apply the projection argument to $|D|^{-1}\int_D\psi_s\,\dd s$, and bound the double integral of the absolute overlaps by extending the Gaussian integral in one variable to $\R^r$.
\end{proof}

The measure here is an ordinary probability weight on the chosen patch; no group-averaging prescription or Jacobian inferred from the gauge algebra is needed. Without a matching lower bound the exponent $r/2$ is not an estimator of orbit dimension. Sections~\ref{sec:phases} and~\ref{sec:volume} supply that lower bound for compact groups.

\subsection{The phases of the overlaps carry the constraint values}\label{sec:phases}

For a compact group the averaging in the proof of Theorem~\ref{thm:resolved} can be carried out over the whole group, and the inequality becomes an identity. This makes it possible to see exactly what the bound with absolute values leaves out.

\begin{proposition}[exact Haar formula]\label{prop:haar}
Let $G$ be compact with normalized Haar measure, and $\psi$ normalized. Then
\begin{equation}\label{eq:haar}
F_\hbar(\psi)=\int_G\langle\psi|U(g)\psi\rangle\,\dd g\ \leq\ \int_G|\langle\psi|U(g)\psi\rangle|\,\dd g,
\end{equation}
with equality on the right if and only if the overlaps are nonnegative almost everywhere. To second order in $\xi$ the overlaps are
\begin{equation}\label{eq:cumulant}
\langle\psi|e^{-i\C_\xi/\hbar}\psi\rangle=\exp\Big(-\tfrac{i}{\hbar}\,c_\xi-\tfrac{1}{2\hbar^2}\Var(\C_\xi)+O(|\xi|^3)\Big).
\end{equation}
\end{proposition}
\begin{proof}
$P_\hbar=\int U(g)\,\dd g$, which is group averaging for a compact group~\cite{abc2005,corichiGA}, and $\|P_\hbar\psi\|^2=\langle\psi|P_\hbar\psi\rangle$. Equation~\eqref{eq:cumulant} is the cumulant expansion of the characteristic function of $\C_\xi$.
\end{proof}

\begin{figure}[t]
\centering
\includegraphics[width=\textwidth]{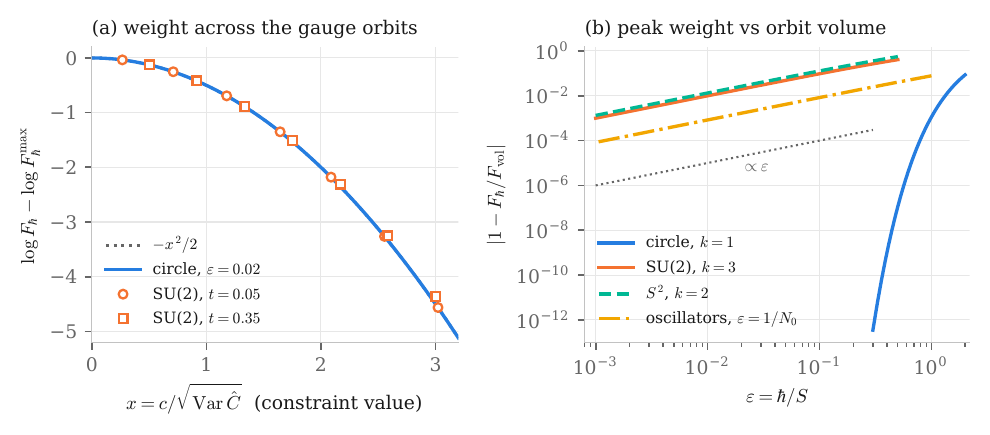}
\caption{The physical weight landscape, computed from exact sums. (a) $\log F_\hbar-\log F_\hbar^{\max}$ along one imaginary gauge line, against the constraint value in units of its standard deviation at the maximum. Circle packets ($\varepsilon=0.02$, line) coincide with the Gaussian law $-x^2/2$ (dotted, hidden under the line); SU(2) heat-kernel packets (markers) follow it, with a small non-Gaussian tail at $t=0.35$. For SU(2) the constraint value is obtained from the exact identity $c=(\hbar/2)\,\dd\log\|\psi\|^2/\dd y$ and the norm formula~\eqref{eq:hallnorm}. (b) Relative deviation of the exact peak weight from the orbit-volume formula~\eqref{eq:volume}. It is of order $\varepsilon$ for SU(2) ($k=3$), for $S^2$ with its U(1) stabilizer ($k=2$) and for two oscillators ($\varepsilon=1/N_0$, $k=1$), and exponentially small for the circle; the dotted line is a slope-one guide.}
\label{fig:landscape}
\end{figure}

So the first inequality in~\eqref{eq:resolved} becomes an equality for Haar-distributed translates, and the second is attained exactly when the overlaps carry no phase. By~\eqref{eq:cumulant} their phases are the constraint values, so the bound with absolute values is attained on $\Sigma$, to leading order. Off $\Sigma$ the phases interfere: inserting~\eqref{eq:cumulant} into~\eqref{eq:haar} and integrating the Gaussian gives the factor $\exp(-\tfrac12c^{\mathsf T}\Cov^{-1}c)$ of~\eqref{eq:gauss}. The circle packets of Sec.~\ref{sec:torus} show this: the ratio of the two sides of~\eqref{eq:haar} agrees with $\exp(-c^2/2\Var)$ to six digits at $c/\sqrt{\Var}=0.6$ and $1.2$. Figure~\ref{fig:landscape}(a) shows the resulting landscape across the gauge orbits for circle and SU(2) packets.

\subsection{The orbit-volume formula}\label{sec:volume}

The same Gaussian integral at $c=0$ gives the peak value of the landscape. Let $\Vol_g(G\cdot[\psi])$ be the volume of the quantum gauge orbit of $[\psi]$ in the metric $g_\hbar$ of~\eqref{eq:kahler}. Its tangent metric at $[\psi]$ is the Gram matrix $\mathsf G$ of~\eqref{eq:gram} on a complement of the stabilizer algebra.

\begin{proposition}[orbit-volume formula]\label{prop:volume}
Let $[\psi]$ lie on $\mu^{-1}(0)$ and have a $k$-dimensional gauge orbit. In each of the following cases,
\begin{equation}\label{eq:volume}
F_\hbar(\psi)=\frac{(4\pi\hbar)^{k/2}}{\Vol_g(G\cdot[\psi])}\,\big(1+O(\hbar)\big).
\end{equation}
\begin{enumerate}[label=(\arabic*)]
\item Linear actions with coherent states centered on $\Sigma$, including nontrivial stabilizers. This is~\eqref{eq:laplace} rewritten: the coherent family is isometric for $g=2|\dd z|^2$, so $\Vol_g=2^{k/2}(\vol(G)/\vol(G_z))\sqrt{\det M_\perp}$.
\item The heat-kernel packets of Secs.~\ref{sec:torus}, \ref{sec:su2} and~\ref{sec:sphere} on $T^r$, SU(2) and $S^2$, and the product packets of Sec.~\ref{sec:structure}. Here both sides of~\eqref{eq:volume} are explicit sums over representations, and the Euler--Maclaurin formula gives their ratio as $1+O(\varepsilon)$ with $\varepsilon=\hbar/S$ (Fig.~\ref{fig:landscape}(b)).
\end{enumerate}
\end{proposition}

The formula has a simple reading. A calibrated packet occupies one cell of $g$-volume $(4\pi\hbar)^{k/2}$ along its gauge orbit, and its physical weight is one over the number of cells. Case~(1) is the Bargmann--Fock counterpart of the diagonal asymptotics of invariant Szeg\H{o} and Bergman kernels on the zero level set, where the reciprocal of the orbit volume (the effective potential of~\cite{burnsguillemin}) appears in the same way~\cite{paoletti2005,paoletti2008,mazhang}. What is new here is its reading as a statement about the physical weight, its extension to the heat-kernel packets of case~(2), and its relation to the resolution bound of Theorem~\ref{thm:resolved}. In case (1) it also identifies the first correction to the reduced K\"ahler potential: on $\Sigma$, $\hbar\log F_\hbar=-\hbar\log\Vol_g(G\cdot z)+\mathrm{const}+O(\hbar^2)$.

A special case of~\eqref{eq:volume} was already implicit in~\cite{corichi2008}. For a single oscillator constrained to a fixed energy, the probability that a coherent state on the classical orbit is physical was found there to behave as $(2\pi n)^{-1/2}$ for large excitation number $n$. This is~\eqref{eq:volume} with $k=1$: the gauge orbit is the classical circle, and about $(2\pi n)^{1/2}$ Planck cells are needed to cover it. The two-oscillator example of Sec.~\ref{sec:linear} is the same statement for the Hopf circles. Combined with Theorem~\ref{thm:resolved}, the orbit-volume formula says that calibrated packets are as physical as a state localized along a gauge orbit can be.

\begin{conjecture}[intrinsic form]\label{conj:intrinsic}
Let $G$ be compact and $[\psi]\in\mu^{-1}(0)$. Suppose the overlaps $\langle\psi|U(g)\psi\rangle$ are negligible outside an $O(\sqrt\hbar)$ neighborhood of the stabilizer, and the cumulants of every $\C_\xi$ scale as $\kappa_n=O(\hbar^{n-1})$. Then~\eqref{eq:volume} holds, and off $\mu^{-1}(0)$ it holds with the extra factor $\exp(-\tfrac12c^{\mathsf T}\Cov^{-1}c)$.
\end{conjecture}

The heuristic is~\eqref{eq:haar} with~\eqref{eq:cumulant}. For a free orbit, $\dd g=\dd\xi/\vol(G)$ near the identity, and the Gaussian integral gives $(4\pi\hbar)^{k/2}/(\vol(G)\sqrt{\det\mathsf G_{ab}})$; with a stabilizer $G_\psi$ the integral along $G_\psi$ contributes a factor $\vol(G_\psi)$. In both cases the denominator is $\Vol_g$, because $U$ acts by isometries. The first hypothesis is the resolution hypothesis of Theorem~\ref{thm:resolved}; the second is what coherent and heat-kernel packets satisfy. The statement is intrinsic: it needs $\Pphy$, the gauge action and the metric of the space of rays, but no packet family.

\paragraph{Calibration and the power of $\hbar$.} For the circle packets, \eqref{eq:volume} holds for any width, in the packets' own metric. With width parameter $\eta=\varepsilon^{2\alpha}$, $F_\hbar=(\eta/\pi)^{1/2}$, and~\eqref{eq:volume} holds up to exponentially small terms in the metric $g_\hbar(\partial_\theta,\partial_\theta)=\hbar/\eta$. That metric converges as $\hbar\to0$ only for $\alpha=1/2$, the K\"ahler-calibrated width. This explains the observation of Sec.~\ref{sec:torus} that the exponent of $F_\hbar$ depends on the shape of the packet. For a calibrated family, $\Vol_g$ tends to the classical orbit volume, so
\begin{equation}\label{eq:limit}
\lim_{\hbar\to0}\ \hbar^{-k/2}F_\hbar(\psi_\hbar)=\frac{(4\pi)^{k/2}}{\Vol_g(\text{classical gauge orbit})}.
\end{equation}
This is the matching lower bound to Corollary~\ref{cor:patch}. In $g$-arclength coordinates, calibrated overlaps are $e^{-d_g^2/4\hbar}$ to leading order, so Corollary~\ref{cor:patch} with $A=1$ gives~\eqref{eq:volume} as an upper bound to the same order. For calibrated packets the exponent of $F_\hbar$ does estimate the dimension $k$ of the gauge orbit.

\section{Examples}\label{sec:examples}

In this section we illustrate the general results in a series of examples, each chosen to probe a different feature of the analysis. The first two have Abelian, or Abelianizable, constraints: commuting momenta on a torus, where every quantity can be computed in closed form, and a model with a structure function, the simplest mechanical analogue of the situation in gravity. The next two have a non-Abelian algebra, acting freely on SU(2) and with a stabilizer on the sphere. We then turn to linear actions, where Theorem~\ref{thm:linear} applies directly; they include a two-dimensional version of the oscillator example of~\cite{corichi2008} and an orbit that is not closed. The last two examples lie outside the hypotheses of Sec.~\ref{sec:projection}: zero in the continuous spectrum, and an anomalous pair with no physical states at all. In every case with a common kernel both readings succeed, and the peak weight agrees with the orbit-volume formula~\eqref{eq:volume}. Table~\ref{tab:examples} summarizes the examples; $\varepsilon=\hbar/S$ for a fixed action scale $S$, $k$ is the gauge-orbit dimension, and the constants refer to the stated packet and metric normalizations.

\begin{table}[t]
\centering\footnotesize
\renewcommand{\arraystretch}{1.25}
\begin{tabularx}{\textwidth}{@{}>{\raggedright\arraybackslash}p{3.3cm}>{\raggedright\arraybackslash}p{2.3cm}c>{\raggedright\arraybackslash}p{2.0cm}>{\raggedright\arraybackslash}p{2.4cm}>{\raggedright\arraybackslash}X@{}}
\toprule
System and preparation & Constraints & $k$ & Peak weight $F_\hbar$ & Ratio to~\eqref{eq:volume} & Recovered \\
\midrule
Zero momenta on $T^r$, periodic Gaussians & $r$, commuting & $r$ & $(\varepsilon/\pi)^{r/2}$ & $1$ up to $e^{-\pi^2/\varepsilon}$ & directions; $\Sigma$ as maximum (Thm~\ref{thm:hall}, $G=U(1)^r$) \\
Curved leaves on $T^3$, adapted packets & 2, structure function & 2 & $\varepsilon/\pi$ & $1$ up to $e^{-\pi^2/\varepsilon}$ & directions; $f'$ from a group commutator \\
Angular generators on SU(2), heat kernel & 3, $\mathfrak{su}(2)$ & 3 & $(2\varepsilon)^{3/2}/(2\sqrt\pi)$ & $1+O(\varepsilon)$ & $\Sigma$ = zero section (Thm~\ref{thm:hall}) \\
Angular generators on $S^2$, heat kernel & 3, $\mathfrak{su}(2)$, rank 2 & 2 & $2\varepsilon$ & $1+O(\varepsilon)$ & leaf dimension from rank \\
Two oscillators at fixed energy & 1, with a character & 1 & $(2\pi N_0)^{-1/2}$ & $0.988$ at $N_0=7$ & Hopf circles (Thm~\ref{thm:linear}) \\
Two vector oscillators, $J=0$ & 3, $\mathfrak{so}(3)$ & 3 & Laplace~\eqref{eq:laplace} & $1.13$ at $\hbar=0.1$, $1.06$ at $\hbar=0.05$ & $\Sigma$, orbits, $\Gamma_\red$ (Thm~\ref{thm:linear}) \\
One vector oscillator, $J=0$ & 3, $\mathfrak{so}(3)$, stabilizer SO(2) & 2 & $\hbar/(2r^2)$ & $1-e^{-2R}$ exactly; $0.982$ at $R=2$ & $\Sigma$ = real vectors (Thm~\ref{thm:linear}) \\
Linear constraint $\hat p_1$ on $L^2(\R^2)$ & 1, zero in continuous spectrum & 1 & rigged: $2(\pi\hbar)^{1/2}$ on $\Sigma$ & not applicable & $\Sigma$ as maximum of the rigged weight \\
Anomalous pair & 2, central term & -- & no physical states & -- & failure detected: $X_{c_1}F_\GB\neq0$ \\
\bottomrule
\end{tabularx}
\caption{Examples. ``Ratio'' compares the exact weight with the orbit-volume formula~\eqref{eq:volume}.}
\label{tab:examples}
\end{table}

\subsection{Commuting constraints on a torus}\label{sec:torus}

We begin with the simplest case. It already displays most features of the general analysis and, in particular, shows why the exponent of the physical weight cannot be read from $\Pphy$ alone. Take $\Gamma=T^*T^r\times\Gamma_\red$ with symplectic form $\sum_j\dd\theta_j\wedge\dd p_j+\omega_\red$ and $C_j=p_j$. On $L^2(T^r,\dd^r\theta/(2\pi)^r)\otimes\Hil_\red$ put $\hat p_j=-i\hbar\partial_{\theta_j}$; the joint kernel consists of constants tensored with $\Hil_\red$. Use one circle packet per direction, with coefficients in the momentum basis
\begin{equation}\label{eq:circle}
\begin{gathered}
\psi_n(\theta,p)=Z_p^{-1/2}\,e^{-\varepsilon(n-p/\hbar)^2/2-in\theta},\\
Z_p=\sum_{n\in\Z}e^{-\varepsilon(n-p/\hbar)^2}=\sqrt{\pi/\varepsilon}\sum_{m\in\Z}e^{-\pi^2m^2/\varepsilon}e^{2\pi imp/\hbar}.
\end{gathered}
\end{equation}
These are periodic Gaussian packets of the familiar circle type~\cite{gonzalezdelolmo}. Normalization of the real amplitudes gives $\langle\psi|\dd\psi\rangle=-i\bar p\,\dd\theta/\hbar$, so the pullback of $\Omega_\hbar$ is $\dd\theta\wedge\dd\bar p$, with $\bar p=p+(\hbar/2\varepsilon)\partial_w\log Z_p$ and $w=p/\hbar$; hence $\bar p\to p$ in $C^1$ on compact sets. The pullback of $g_\hbar$ tends to $S\,\dd\theta^2+\dd p^2/S$, so the family is K\"ahler calibrated. The unitary itself translates the packet, which proves propagation. For the circle character $E$, $\|(E-e^{i\theta})\psi\|=O(\sqrt\varepsilon)$ uniformly in the center, and $\|(\hat p-p)\psi\|=O(S\sqrt\varepsilon)$.

On $\Sigma$ take $p_j=0$ and a normalized spectator $\chi$. All constraint means vanish and each variance is asymptotic to $\hbar S/2$. The normalized joint projection is the constant torus mode tensored with $\chi$, with probability
\begin{equation}\label{eq:torusF}
F_\hbar=Z_0^{-r}=(\varepsilon/\pi)^{r/2}\big(1+O(e^{-\pi^2/\varepsilon})\big).
\end{equation}
This saturates Corollary~\ref{cor:patch}. Yet~\eqref{eq:fields} gives a nonzero limiting squared speed $S$ per circle: weight, residual and speed are distinct quantities.

The projection reading adds two facts. First, \eqref{eq:circle} is the $U(1)^r$ heat-kernel packet of Theorem~\ref{thm:hall} with heat time $\varepsilon/2$. Second, off $\Sigma$ the weight is $F_\hbar=\prod_je^{-\varepsilon w_j^2}/Z_{p_j}$, which is~\eqref{eq:gauss} with $\Var(\hat p_j)=\hbar^2/(2\varepsilon)=\hbar S/2$. So the maximum of $F_\hbar$ over imaginary translates lies exactly on $\Sigma$.

Recovering the circle direction cannot fix the spectral zero. On periodic functions $\hat p+\hbar\beta$ has the same ray motion for every real $\beta$, but its least squared residual is $\hbar^2\,\mathrm{dist}(\beta,\Z)^2$, by its Fourier eigenvalues. This spectral information belongs to the specified operator.

What fixes the exponent in~\eqref{eq:torusF}? Give the Gaussian the shape parameter $\eta=\varepsilon^{2\alpha}$ instead, with $0<\alpha<1$. Both angle and momentum still localize, but $F_\hbar\sim\pi^{-r/2}\varepsilon^{\alpha r}$. Moreover $\PP(K_\hbar)$ is isometric to $\PP(\Hil_\red)$ at fixed $\hbar$ for every $r$, while the leaves have dimension $r$. So the leaf dimension cannot be read from the intrinsic physical K\"ahler space. Section~\ref{sec:volume} resolves the exponent: \eqref{eq:volume} holds for every $\alpha$ in the packets' own metric, and that metric converges only for $\alpha=1/2$.

\subsection{Two constraints with a structure function}\label{sec:structure}

Our second example is designed to have a structure function, and it is the simplest mechanical model of the situation in gravity. It also separates the intrinsic data, the common kernel and the null foliation, from those that depend on the presentation of the constraints. On $T^*T^3$ with angles $x,y,u$ take~\eqref{eq:T3}. The coordinates $v=u-g(x)$, $P_x=p_x+g'(x)p_u$ and $p_v=p_u$ are adapted to the leaves: $p_x\dd x+p_u\dd u=P_x\dd x+p_v\dd v$. On $\Sigma$, $P_x=p_y=0$. The restricted form is $\dd v\wedge\dd p_v$, the reduced space is $T^*S^1$, and a leaf is a two-torus at fixed $(v,p_v)$. In the original coordinates $u=v+g(x)$ and $p_x=-g'(x)p_v$ on $\Sigma$, so even one of the measured momenta varies along a leaf. The fields and their brackets are
\begin{equation}\label{eq:T3fields}
\begin{gathered}
X_1=\partial_x+g'\partial_u-g''p_u\,\partial_{p_x},\qquad X_2=f\,\partial_y-f'p_y\,\partial_{p_x},\\
\{C_1,C_2\}=-\tfrac{f'}{f}\,C_2,\qquad [X_1,X_2]\big|_\Sigma=f'\,\partial_y .
\end{gathered}
\end{equation}
The commuting frame $P_x,p_y$ is globally available, so the structure function is not an intrinsic obstruction to Abelianization; specifying the individual transformations makes the two presentations distinguishable.

\paragraph{Motion.} On $L^2(T^3)$, the volume-preserving base flows $T_1^t(x,y,u)=(x+t,y,u+g(x+t)-g(x))$ and $T_2^t(x,y,u)=(x,y+tf(x),u)$ define self-adjoint generators, with smooth periodic functions as a common core. With $h_t(x)=g(x+t)-g(x)$, conjugating the momenta gives
\begin{equation}\label{eq:T3conj}
U_1(t)^*\hat p_xU_1(t)=\hat p_x-h_t'(x)\,\hat p_u,\qquad U_2(t)^*\hat p_xU_2(t)=\hat p_x-tf'(x)\,\hat p_y .
\end{equation}
Starting at $p_x=-g'(x_0)p_u$, the first motion gives $p_x(t)=-g'(x_0+t)p_u$, an on-shell test of a cotangent response. For product packets~\eqref{eq:circle} centered in a compact momentum set, $\|(\hat p_j-p_j(t))U_i(t)\psi\|\leq C\sqrt\varepsilon$ uniformly on bounded times. Finite words have base maps $(x+a,\,y+b(x),\,u+c(x))$, and the same argument applies. Comparing $T_1^s$ followed by $T_2^t$ with the reverse order changes $y$ by $t[f(x+s)-f(x)]$. The limiting readings of $e^{iy}$ give that difference as a relative phase; dividing by $st$ and letting $s,t\to0$ recovers $f'(x)$. Replacing $C_2$ by $p_y$ preserves the common kernel and the characteristic foliation, but removes this response.

\paragraph{Projection.} In adapted coordinates the operators are $-i\hbar\partial_x$ at fixed $v$ and $f(x)\hat p_y$. Their common kernel is $L^2(S^1_v)$, embedded as functions independent of $x$ and $y$. The projector is the conditional mean
\begin{equation}\label{eq:T3proj}
(P\psi)(x,y,u)=\int_{T^2}\psi\big(\xi,\eta,u-g(x)+g(\xi)\big)\,\frac{\dd\xi\,\dd\eta}{(2\pi)^2}.
\end{equation}
For zero-momentum packets in the adapted $x,y$ variables tensored with any normalized $\chi(v)$, the normalized projection is $\chi(v)$ and the weight is $Z_0^{-2}\sim\varepsilon/\pi$. The operators $e^{iv}$ and $\hat p_v$ preserve the physical sector and quantize the surviving canonical pair. In adapted coordinates $-i\hbar\partial_x$ and $\hat p_y$ generate a $T^2$ action, so Theorem~\ref{thm:hall} with $G=U(1)^2$ applies to these packets. There is no group in the $(\C_1,\C_2)$ presentation; still, $F_\hbar$ is the same function in both presentations (Sec.~\ref{sec:dirac}), and its maxima over imaginary translates lie on $\langle\hat P_x\rangle=\langle\hat p_y\rangle=0$.

\subsection{A free orbit on SU(2)}\label{sec:su2}

Next we consider a non-Abelian gauge group acting freely, the simplest case in which the constraints do not commute and the gauge orbit is a non-Abelian group manifold. Let $G=SU(2)$, $[\tau_a,\tau_b]=\epsilon_{ab}{}^c\tau_c$, and $Y_a(g)=g\tau_a$. On $T^*G$ put $C_a(g,p)=p(Y_a(g))$. These left-invariant fields generate right multiplication, and $\{p(Y),p(Z)\}=-p([Y,Z])$. The joint zero set is the zero section, on which the characteristic fields restrict to $Y_a$. On $L^2(G,\dd g)$ with normalized Haar measure, $\C_a=-i\hbar Y_a$ satisfy $[\C_a,\C_b]/(i\hbar)=-\epsilon_{ab}{}^c\C_c$. Connectedness and transitivity make the common kernel the constants; no individual generator imposes this condition. The normalized packet at $g_0$ and its exact weight are
\begin{equation}\label{eq:su2}
\begin{gathered}
\psi_{t,g_0}(g)=\frac{H_t(g_0^{-1}g)}{\sqrt{H_{2t}(e)}},\qquad F_\hbar=\frac1{H_{2t}(e)},\\
H_s(e)=\sum_{j=0,\frac12,\dots}(2j+1)^2e^{-sj(j+1)}=e^{s/4}\sum_{n\geq1}n^2e^{-sn^2/4}\sim2\sqrt\pi\,s^{-3/2}.
\end{gathered}
\end{equation}
With $t=\varepsilon$ this gives $F_\hbar\sim(2t)^{3/2}/(2\sqrt\pi)$. By centrality the $a$-th constraint moves the center exactly to $g_0e^{s\tau_a}$. The group commutator $e^{s\tau_a}e^{u\tau_b}e^{-s\tau_a}e^{-u\tau_b}$ moves the center by $su\,g_0[\tau_a,\tau_b]+o(su)$, and bounded matrix entries detect the third direction. The means vanish, and each variance is asymptotic to $\hbar^2/(4t)$, so all three residuals vanish while the three orbit speeds have nonzero limits. Propagation is claimed on the zero section, not at arbitrary nonzero covectors.

Theorem~\ref{thm:hall} places these packets inside the complexified family $\psi_z$: they are its members with $Y=0$, the maximizers of $F_\hbar$. Write $\Lambda(s)=-\dd\log H_s(e)/\dd s$. The orbit metric is $g(Y_a,Y_b)=\delta_{ab}\,(2\hbar/3)\,\Lambda(2t)$, and the metric with $|\tau_a|=1$ gives SU(2) the volume $16\pi^2$. So~\eqref{eq:volume} reads $F_\hbar=(6\pi)^{3/2}(16\pi^2)^{-1}\Lambda(2t)^{-3/2}$, which agrees with~\eqref{eq:su2} to $1+O(t)$.

\subsection{The angular generators on the sphere}\label{sec:sphere}

The sphere shows what happens when the gauge action has a stabilizer: there are three constraints, but the gauge orbits are two dimensional. Use $S^2=SU(2)/U(1)$, with the unit round metric and area measure normalized to one. The constraints on $T^*S^2$ are the components of $L=n\times\pi$, where $n^2=1$ and $n\cdot\pi=0$. Since $n\cdot L=0$ identically, the three constraints have only two independent differentials on the zero section. Their fields rotate $n$; rotation about $n$ fixes it. The characteristic leaf has dimension two, and the common zero mode is the constant spherical harmonic. Use the round heat kernel $H_t(n,n_0)$ divided by $\sqrt{H_{2t}(n_0,n_0)}$. It rotates with its center, and a rotation about $n_0$ fixes the entire packet even before projection. Homogeneity and the degree-$\ell$ multiplicity give
\begin{equation}\label{eq:sphere}
H_s(n_0,n_0)=\sum_{\ell\geq0}(2\ell+1)e^{-s\ell(\ell+1)}\sim s^{-1},\qquad F_\hbar\sim2t .
\end{equation}
With center at the north pole, $\hat L_z$ annihilates the packet, and the two transverse variances are each asymptotic to $\hbar^2/(4t)$. A commutator of transverse rotations generates, to leading order at the pole, a stabilizer rotation; a vanishing displacement of the center does not imply a vanishing operator bracket. In Corollary~\ref{cor:patch} one must use a transverse two-dimensional patch: three rotation parameters do not yield three distinguishable directions. Write $\Lambda(s)=-\dd\log H_s(n_0,n_0)/\dd s$. In~\eqref{eq:volume} the orbit is the sphere with $\hbar\Lambda(2t)$ times the round metric, so~\eqref{eq:volume} reads $F_\hbar=1/\Lambda(2t)$, which agrees with~\eqref{eq:sphere} to $1+O(t)$.

\subsection{Linear actions}\label{sec:linear}

We now turn to linear actions, for which Theorem~\ref{thm:linear} applies directly and all statements hold at every $\hbar$. The first example is a two-dimensional version of the oscillator example of~\cite{corichi2008}; the last one shows what happens on an orbit that is not closed.

\paragraph{Two oscillators at fixed energy.} Here $\C=\hbar\nu(\hat N+1)-E$ with $E=\hbar\nu(N_0+1)$, and $U(1)$ acts by $z\mapsto e^{-i\nu t}z$ on $\Cx^2$. The constant term is a character, so $P\cs{\lambda z}=\lambda^{N_0}P\cs z$ for $\lambda\in\Cx^*$: the physical ray is invariant, though the vector is not. With $R=|z|^2/\hbar$, $F_\hbar=e^{-R}R^{N_0}/N_0!$. On each $\Cx^*$-orbit, $F_\hbar$ is maximal at $R=N_0$, which is where $\langle z|\C|z\rangle=\hbar\nu(R-N_0)=0$; the classical surface $\nu|z|^2=E$ is shifted by the zero-point term $\hbar\nu$. The fibers of $\pi$ are the $\Cx^*$-orbits and the gauge orbits are the Hopf circles. The physical form $N_0\hbar\,\omega_{\mathrm{FS}}$ tends to $\omega_\red$; this limit is derived in~\cite{corichiMC}, where the master evolution of the same coherent packets is also obtained in closed form. The orbit is a circle of $g$-length $2\pi\sqrt{2\hbar R}$, so~\eqref{eq:volume} gives $(2\pi N_0)^{-1/2}$ exactly.

\paragraph{Two vector oscillators, $J=0$.} Let $z=(z_1,z_2)\in\Cx^3\oplus\Cx^3$, with SO(3) rotating both vectors. The constraints $\hat J_a$ satisfy $[\hat J_a,\hat J_b]=i\hbar\epsilon_{abc}\hat J_c$, and their coherent expectations are the classical angular momentum $q_1\times p_1+q_2\times p_2$. The complexified group is $SO(3,\Cx)$, whose invariants are the bilinears $s_{kl}=z_k\cdot z_l$, without complex conjugation. For two vectors they are algebraically independent, so $\Gamma_\red=\Sigma/SO(3)$ has real dimension $12-3-3=6$. Theorem~\ref{thm:linear} then gives three things: $\pi(z)$ depends only on $(s_{11},s_{12},s_{22})$ on the polystable set; along each $SO(3,\Cx)$-orbit, $F_\hbar$ is maximal exactly where the total angular momentum vanishes; and the gauge orbits are the rigid rotations of the pair.

\paragraph{One vector oscillator: a stabilizer.} For a real vector $z\in\R^3\subset\Cx^3$ of length $r$, $J=0$ and the stabilizer is SO(2). Haar averaging gives $(z|P\cs z=\langle e^{(r^2/\hbar)\cos\theta}\rangle=\sinh R/R$ with $R=r^2/\hbar$. Hence $F_\hbar=e^{-R}\sinh R/R\to\hbar/(2r^2)$. Formula~\eqref{eq:laplace}, with $\vol(SO(3))=8\pi^2$, $\vol(SO(2))=2\pi$ and $\det M_\perp=r^4$, gives the same limit; so does~\eqref{eq:volume}, since the orbit is a sphere of $g$-area $8\pi r^2$. This is the linear counterpart of the sphere example of Sec.~\ref{sec:sphere}, and here Theorem~\ref{thm:linear} is proved.

\paragraph{An unstable orbit.} Circular polarization, $z_1=a(1,i,0)/\sqrt2$ and $z_2=0$, has $J=|a|^2\hat e_z$. All invariants vanish, and complex rotations shrink the state towards the vacuum. So $\pi(z)$ is the vacuum ray, and $F_\hbar$ tends to 1 along the orbit without a maximum. This is the quantum form of the fact that the angular momentum of a single circularly polarized vector cannot be rotated away.

\subsection{Zero in the continuous spectrum}\label{sec:rigging}

Our last two examples lie outside the hypotheses of Sec.~\ref{sec:projection}. In the first, zero lies in the continuous spectrum of the constraint, the situation of loop quantum cosmology with a massless scalar field~\cite{aps2006} and, more generally, of noncompact gauge groups. Take $\C=\hat p_1$ on $L^2(\R^2)$. The group $\R$ is noncompact and $K_\hbar=\{0\}$. Replace $F_\hbar$ by the rigged weight $F_R(\psi)=\int\dd s\,\langle\psi|e^{is\hat p_1/\hbar}\psi\rangle/\|\psi\|^2$, the group-averaging pairing~\cite{abc2005} divided by the norm. Since $\delta(\hat p_1)e^{\tau\hat p_1/\hbar}=\delta(\hat p_1)$, the numerator is constant along imaginary flows, and Proposition~\ref{prop:identities} and Corollary~\ref{cor:line} hold for $F_R$ as stated. For coherent states of width $(\hbar/2)^{1/2}$,
\begin{equation}\label{eq:rigged}
F_R(q,p)=2(\pi\hbar)^{1/2}\,e^{-p_1^2/\hbar}.
\end{equation}
It is maximal exactly on $\Sigma=\{p_1=0\}$, and Gaussian in the constraint value with $\Var(\hat p_1)=\hbar/2$, as~\eqref{eq:gauss} requires. The physical image is $[\ket{q_2,p_2}\otimes\ket{p_1=0}]$. So the fibers are the $(q_1,p_1)$-planes, the complexified translations, and the gauge orbits are the lines $p_1=0$ at fixed $(q_2,p_2)$. The rigged weight is not a probability and the orbit has infinite volume, so~\eqref{eq:volume} does not apply.

\subsection{An anomalous pair}\label{sec:anomaly}

Finally, we consider a case with no physical states at all, and ask whether the physical weight can detect the failure. Take $\C_1=\hat p_1$ and $\C_2=\hat p_2-\hbar\kappa\hat q_1$, so $[\C_1,\C_2]=i\hbar^2\kappa$, case~(2) of Proposition~\ref{prop:classification}. This pair is analyzed with master constraints in~\cite{corichiAnom}, and its Gupta--Bleuler states minimize the anomaly bound of~\cite{corichi2008}; here we only record the master-free diagnostic. There is no common kernel, and $\mu^{-1}(0)$ is not coisotropic, since $\{c_1,c_2\}=\hbar\kappa$. The natural substitute for $K_\hbar$ is the Gupta--Bleuler space $\ker(\C_1+i\C_2)$, the lowest Landau level for $\kappa>0$, with weight $F_\GB$. Its projector does not annihilate the $\C_a$, so Proposition~\ref{prop:identities} fails, and visibly so: the flow of $c_1$ shifts $c_2$ at the rate $\hbar\kappa$, so it moves states across the level sets of $F_\GB$. The pair is unitarily equivalent to $\sqrt{\hbar\kappa}$ times a canonical pair $(\hat X,\hat P)$; at $\hbar=\kappa=1$, for the coherent state $\alpha=0.4+0.3i$ of that pair, $X_{c_1}F_\GB=0.33$. An anomaly thus shows up, without a master function, as real gauge fields that are not tangent to the level sets of the physical weight.

\section{Discussion}\label{sec:discussion}

Let us summarize. We began with a question raised by our earlier work~\cite{corichi2008}: if the space of physical quantum states carries no gauge directions, where does the quantum theory encode the classical gauge orbits? The answer we have found is that the classical gauge geometry is not in the physical state space, but in the relation between the physical and the kinematical states. It is in how kinematical packets move under the constraints, and in how they project onto $\Pphy$. Only the classical description by packet labels requires a quotient.

On the exact physical submanifold each individual constraint generates the zero vector, so searching its tangent spaces for a remnant of a classical gauge orbit cannot succeed. On the kinematical space the same generator moves suitably localized preparations, and their limiting probe readings supply velocities at points of $\Sigma$. When the correspondence, tangency and rank tests of Sec.~\ref{sec:motion} hold, those velocities exhaust the characteristic distribution. The projection reading supplies what motion does not: which labels lie on $\Sigma$. For a compact gauge group, within each projection fiber the weight $F_\hbar$ is maximal exactly on one real gauge orbit, and Gaussian in the constraint values away from it. For linear and heat-kernel families this holds exactly, at every $\hbar$, and it also yields the reduced space with its K\"ahler potential.

The structure-function model makes both sides explicit. Along one gauge direction the angle and its conjugate momentum in the original chart change, while $v=u-g(x)$ and $p_v$ stay fixed. Exact imposition gives a wavefunction of $v$ alone. The two coordinates needed to locate a point on a gauge leaf have a meaning for kinematical preparations and are already absent from the physical wavefunction; nothing further should be identified among its rays. With three angular constraints it is the rank of the observed action that counts: a rotation can fix the center of a packet on the sphere even though its generator acts nontrivially elsewhere.

Let us comment on the three lessons that we find most important.

\emph{The imaginary gauge directions matter.} In the symplectic description of classical reduction the gradients $\nabla C_a$ play no role; only the Hamiltonian vector fields do. In the quantum theory both are on the same footing: the complex structure of the space of rays turns each into the other, and the imaginary flows are generated by the same operators as the real ones. The fibers of Dirac projection contain the complexified gauge orbits, and it is along the imaginary directions that the physical weight detects the constraint surface. In this sense Dirac quantization is closer to the holomorphic description of reduction recalled in Sec.~\ref{sec:classical} than to the symplectic one, and Theorem~\ref{thm:KN} is the precise statement of this.

\emph{The physical weight is the intrinsic object.} The zero set of the constraint expectation values, its coisotropy at finite $\hbar$ and the probe responses all depend on how the constraints are written. The physical projector and the weight $F_\hbar$ do not. Proposition~\ref{prop:identities} shows that $F_\hbar$ obeys exact identities in every presentation, with or without closure of the constraint algebra, and the projection reading is built on it. For theories with structure functions, and in particular for gravity, this is the feature we expect to be most useful.

\emph{Gauge localization has a price.} A kinematical state that resolves a gauge orbit has a small probability of being physical, and for calibrated packets that probability is one over the number of Planck cells that cover the orbit. When the gauge orbits are noncompact, or infinite dimensional as in field theory, this probability vanishes, which is the familiar fact that physical states are then not normalizable in the kinematical Hilbert space. The rigged weight of Sec.~\ref{sec:rigging} indicates how the analysis should be modified in that case.

Table~\ref{tab:intrinsic} separates what is intrinsic from what is chosen. The recovery is conditional, and the conditions are explicit. Different classical presentations can have isometric physical ray spaces (Sec.~\ref{sec:torus}). The packet family is the one ingredient that cannot be derived from $\Pphy$, so propagation checks and calibration are essential. What the projection reading shows is that, once a calibrated family is covariant under the complexified gauge group, nothing else is needed.

\begin{table}[t]
\centering\small
\renewcommand{\arraystretch}{1.2}
\begin{tabularx}{0.92\textwidth}{@{}>{\raggedright\arraybackslash}p{6.4cm}>{\raggedright\arraybackslash}X@{}}
\toprule
Object & Depends on \\
\midrule
$\Pphy$, $P_\hbar$ and the weight $F_\hbar$ & the constraints as a system: the same for any presentation with the same common kernel \\
Quantum gauge orbits and their $g_\hbar$-volume & the gauge action on the space of rays \\
$\mu^{-1}(0)$, its coisotropy at finite $\hbar$, the responses $R_a^\mu$ & the presentation of the constraints (Secs.~\ref{sec:dirac} and~\ref{sec:structure}) \\
Packet family, probes, calibration & an operational choice, not determined by $\Pphy$ \\
\bottomrule
\end{tabularx}
\caption{What the recovery uses, and what each ingredient depends on.}
\label{tab:intrinsic}
\end{table}

\paragraph{Semiclassical states and quantum gravity.} Our results bear on the construction of semiclassical states for constrained systems~\cite{abc2005}. There, physical semiclassical states are obtained by group averaging kinematical coherent states. The present analysis describes what is lost in that step and how it can be recovered: the location along the gauge orbit is lost, but the kinematical packets, their complexified orbits and their physical weights together determine $\Sigma$, the gauge orbits and the reduced K\"ahler structure. In loop quantum gravity the heat-kernel packets of Theorem~\ref{thm:hall} are the building blocks of the complexifier coherent states~\cite{thiemannGCS,thiemannwinkler}, and the Gauss constraint acts on them by translations at the vertices; open problem~\ref{op:closure} below is a first step in that direction. In loop quantum cosmology with a massless scalar field~\cite{aps2006}, zero lies in the continuous spectrum of the constraint, and the rigged weight of Sec.~\ref{sec:rigging} is the relevant object. Finally, the Hamiltonian constraint of general relativity closes only with structure functions. There, Proposition~\ref{prop:identities} and the presentation independence of $F_\hbar$ are the parts of the projection reading that survive without a group, and open problem~\ref{op:gravity} asks how much more can be said.

\paragraph{Open problems.}
\begin{enumerate}[label=(\roman*)]
\item \emph{Full proofs with domains:} the holomorphic extension of $U$ to entire vectors used in Theorem~\ref{thm:KN} for unbounded generators, the Laplace remainder in~\eqref{eq:laplace}, and the response identity for Berezin--Toeplitz probes on calibrated families (Sec.~\ref{sec:motion}).
\item \emph{The intrinsic orbit-volume law:} prove Conjecture~\ref{conj:intrinsic}, with uniform error terms and nontrivial stabilizers. This would make the exponent of $F_\hbar$ a certified estimator of the gauge-orbit dimension.
\item \emph{Nonlinear compact actions.} For a compact group acting on a compact K\"ahler manifold, with $\hbar$ the inverse power of the prequantum line bundle, the tool is the asymptotic expansion of invariant Bergman kernels~\cite{mazhang}. We conjecture that Theorem~\ref{thm:linear} holds with $O(\hbar)$ errors.
\item \emph{The sphere.} Coherent states on $T^*S^2$ with a complexified SO(3) action~\cite{hallmitchell} should place the packets of Sec.~\ref{sec:sphere} under Theorem~\ref{thm:KN}; the one-vector oscillator of Sec.~\ref{sec:linear} is the linear model.
\item\label{op:gravity} \emph{Structure functions and gravity.} Proposition~\ref{prop:identities} needs no closure, and $F_\hbar$ does not depend on the presentation. In the $T^3$ model a commuting frame exists globally; in gravity it does not. Is there a set of expectation-value constraints, possibly including moments as in effective constraints~\cite{bojowald}, whose zero set is coisotropic and is still characterized by stationarity of $F_\hbar$?
\item \emph{Zero in the continuous spectrum.} The rigged weight works for $\R^k$ (Sec.~\ref{sec:rigging}). $SL(2,\R)$, other non-amenable groups, and a Gauss law with a Schwinger term remain open.
\item \emph{Rank changes.} Theorem~\ref{thm:linear} describes the polystable set, and the unstable orbit of Sec.~\ref{sec:linear} shows where $F_\hbar$ has no maximum. The stratified picture~\cite{sjamaarlerman} should follow, but the local regular argument of Sec.~\ref{sec:motion} does not settle it.
\item \emph{Second class.} The holomorphic combination $\C_1+i\C_2$ annihilates the Gupta--Bleuler space. This suggests a half-sized Kempf--Ness theorem, which would connect to the anomaly analysis of~\cite{corichiAnom}.
\item\label{op:closure} \emph{SU(2) closure.} In the spinor formulation the action is linear on $(\Cx^2)^N$, so Theorem~\ref{thm:linear} applies: closed polyhedra are the coherent states of maximal physical weight on their $SL(2,\Cx)$-orbits. This must be compared carefully with~\cite{conradyfreidel,bds} before any novelty is claimed for this case.
\end{enumerate}

We hope that the picture developed here, in which the classical gauge structure is recovered from the way kinematical states project onto the physical ones, will prove useful in the semiclassical analysis of constrained quantum theories and, in particular, of quantum gravity.

\section*{Acknowledgments}
I thank Steph for discussions and for an earlier draft that developed the motion reading, the resolution bound and several of the examples, and Fide for support. This work received no external funding. 
During the preparation of this manuscript the author used Claude (Anthropic; Claude Opus~5.5) for assistance with drafting and editing the text, s and generate the figure. The author has reviewed and edited the output and takes full responsibility for the content of this publication.

\section*{Author declarations} There are no declarations
\paragraph{Conflict of interest.} The author has no conflicts to disclose.


\appendix
\section{Numerical checks}\label{app:checks}

Every new quantitative claim was checked numerically with scripts that use only \texttt{numpy} and \texttt{scipy} (Python~3.13); they are provided as supplementary material. SU(2) Haar integrals use Euler angles on a $56\times64\times128$ grid (Gauss--Legendre in $\cos\theta$), and SO(3) integrals a $40\times48\times48$ grid. Table~\ref{tab:checks} lists the checks and their outcomes; all pass. Further checks of Theorem~\ref{thm:KN} for SU(2) on two spin-1 particles, of the oscillator and two-vector examples, of circular polarization, and of the reduced K\"ahler potential of Theorem~\ref{thm:linear}(5) for a torus action (to $4\times10^{-8}$) were carried out with the same tools.

\begin{table}[ht]
\centering\footnotesize
\renewcommand{\arraystretch}{1.18}
\begin{tabularx}{\textwidth}{@{}>{\raggedright\arraybackslash}X>{\raggedright\arraybackslash}p{6.6cm}@{}}
\toprule
Claim and system & Result \\
\midrule
Classical relation~\eqref{eq:classKN} and the moment-map convention, random generator on $\Cx^5$ & errors $6\times10^{-9}$ and $3\times10^{-10}$ (finite differences); $\mu=0$ at the minimum of the norm for U(1) on $\Cx^2$ \\
Proposition~\ref{prop:identities}: three random noncommuting constraints on $\Cx^{14}$, common kernel of dimension 3, $\hbar=0.6$ & errors $8\times10^{-12}$ (imaginary), below $10^{-15}$ (real); $\|[\C_1,\C_2]\|=2.26$ \\
$P_\hbar W=P_\hbar$ for a word mixing real and imaginary flows & relative error $2\times10^{-16}$ with $\|W\|=110$ \\
Theorem~\ref{thm:hall}(1): $P\psi_z=1$ and $U(h)\psi_z=\psi_{zh}$ for complex $z,h$ (SU(2)) & $1.000000000000$; error $3\times10^{-15}$ \\
Norm formula~\eqref{eq:hallnorm} & quadrature and formula agree to 10 digits \\
Theorem~\ref{thm:hall}(3): minimum of $\|\psi_z\|^2$ at $Y=0$ & rises from $7.21$ to $126.4$ as $|y|$ goes from 0 to $1.5$ ($t=0.35$) \\
Theorem~\ref{thm:hall}(4): $\dd\log F/\dd\tau=-(2/\hbar)\langle\C_a\rangle$, $t=0.35$ & error $5\times10^{-11}$ \\
Gaussian law~\eqref{eq:gauss} on SU(2) & ratio $0.999999$ at $y=0.01$ \\
Orbit-volume formula~\eqref{eq:volume}: circle & ratio $1.000000$ at $\varepsilon=0.5,0.1,0.01$ \\
\eqref{eq:volume}: SU(2) & ratio $0.816,\,0.951,\,0.990,\,0.9975$ at $t=0.2,0.05,0.01,0.0025$ \\
\eqref{eq:volume}: $S^2$ & ratio $0.753,\,0.935,\,0.987,\,0.9967$ at the same $t$ \\
\eqref{eq:volume}: two oscillators & ratio $0.988,\,0.998,\,0.9998$ at $N_0=7,50,400$ \\
One vector oscillator: $F_\hbar=e^{-R}\sinh R/R$ and~\eqref{eq:laplace} with a stabilizer & quadrature exact to $2\times10^{-15}$; ratio to $\hbar/(2r^2)$ is $0.982$ at $R=2$ \\
Proposition~\ref{prop:haar}: overlap phases give $\exp(-c^2/2\Var)$ (circle) & $0.835270$ and $0.486752$, both to six digits \\
\eqref{eq:volume} for circle packets of width $\varepsilon^\alpha$ in their own metric & ratio $1$ to $10^{-6}$; exponent of $F$ equals $\alpha=0.25,0.50,0.75$ \\
Section~\ref{sec:dirac}: presentation dependence on $T^3$ & $c_1=c_2=0$ with $\{c_1,c_2\}=-0.4167=-\hbar/(2\rho)$ at $\rho=1.2$; second state $\langle\hat p_y\rangle=0.167=\hbar/\rho$ at $\rho=6$ \\
Commutator~\eqref{eq:T3}, spectral derivatives on $T^3$ & relative error $3\times10^{-14}$ \\
Rigged weight~\eqref{eq:rigged} and its identity & errors $7\times10^{-12}$ and $2\times10^{-11}$ \\
Laplace formula~\eqref{eq:laplace}, two vector oscillators & ratio $1.13$ at $\hbar=0.1$, $1.06$ at $\hbar=0.05$ \\
Anomalous pair, $\hbar=\kappa=1$ & $X_{c_1}F_\GB=0.33$ \\
\bottomrule
\end{tabularx}
\caption{Numerical checks of the main statements.}
\label{tab:checks}
\end{table}

\FloatBarrier

\end{document}